\documentclass[table]{article}

\usepackage{arxiv}

\usepackage{natbib}
\usepackage{xcolor}
\usepackage{multirow}
\usepackage{booktabs}
\usepackage{subcaption}

\usepackage{highlight}
\usepackage{xfp}
\usepackage{enumitem}
\usepackage{tabularx}
\usepackage{amsmath}

\usepackage[utf8]{inputenc} 
\usepackage[T1]{fontenc}    
\usepackage{hyperref}       
\usepackage{url}            
\usepackage{booktabs}       
\usepackage{amsfonts}       
\usepackage{nicefrac}       
\usepackage{microtype}      
\usepackage{lipsum}
\usepackage{graphicx}
\graphicspath{ {./images/} }

\title{What Drives Cross-lingual Ranking? Retrieval Approaches with Multilingual Language Models} 

\author{
 Roksana Goworek\footnotemark[1]\\
  The Alan Turing Institute\\
  Queen Mary University of London\\
   \And
  Olivia Macmillan-Scott\footnotemark[1]\\
  The Alan Turing Institute \\
  University College London \\
  \And
  Eda B. Özyiğit  \\
  The Alan Turing Institute\\
}

\begin{document}
\maketitle
\vspace{-3em}
\begin{center}

\texttt{\{rgowerek, omacmillan-scott, eozyigit\}@turing.ac.uk}
\end{center}
\vspace*{2em}

\begingroup\def\thefootnote{*}\footnotetext{Work done during an internship at The Alan Turing Institute.}\endgroup

\begin{abstract}

Cross-lingual information retrieval (CLIR) enables access to multilingual knowledge but remains challenging due to disparities in resources, scripts, and weak cross-lingual semantic alignment in embedding models. Existing pipelines often rely on translation and monolingual retrieval heuristics, which add computational overhead and noise, degrading performance. This work systematically evaluates four intervention types, namely document translation, multilingual dense retrieval with pretrained encoders, contrastive learning at word, phrase, and query–document levels, and cross-encoder re-ranking, across three benchmark datasets. We find that dense retrieval models trained specifically for CLIR consistently outperform lexical matching methods and derive little benefit from document translation. Contrastive learning mitigates language biases and yields substantial improvements for encoders with weak initial alignment, and re-ranking can be effective, but depends on the quality of the cross-encoder training data. Although high-resource languages still dominate overall performance, gains over lexical and document-translated baselines are most pronounced for low-resource and cross-script pairs. These findings indicate that cross-lingual search systems should prioritise semantic multilingual embeddings and targeted learning-based alignment over translation-based pipelines, particularly for cross-script and under-resourced languages.

\end{abstract}

\begin{figure}[h]
    \centering
    \includegraphics[width=0.9\linewidth]{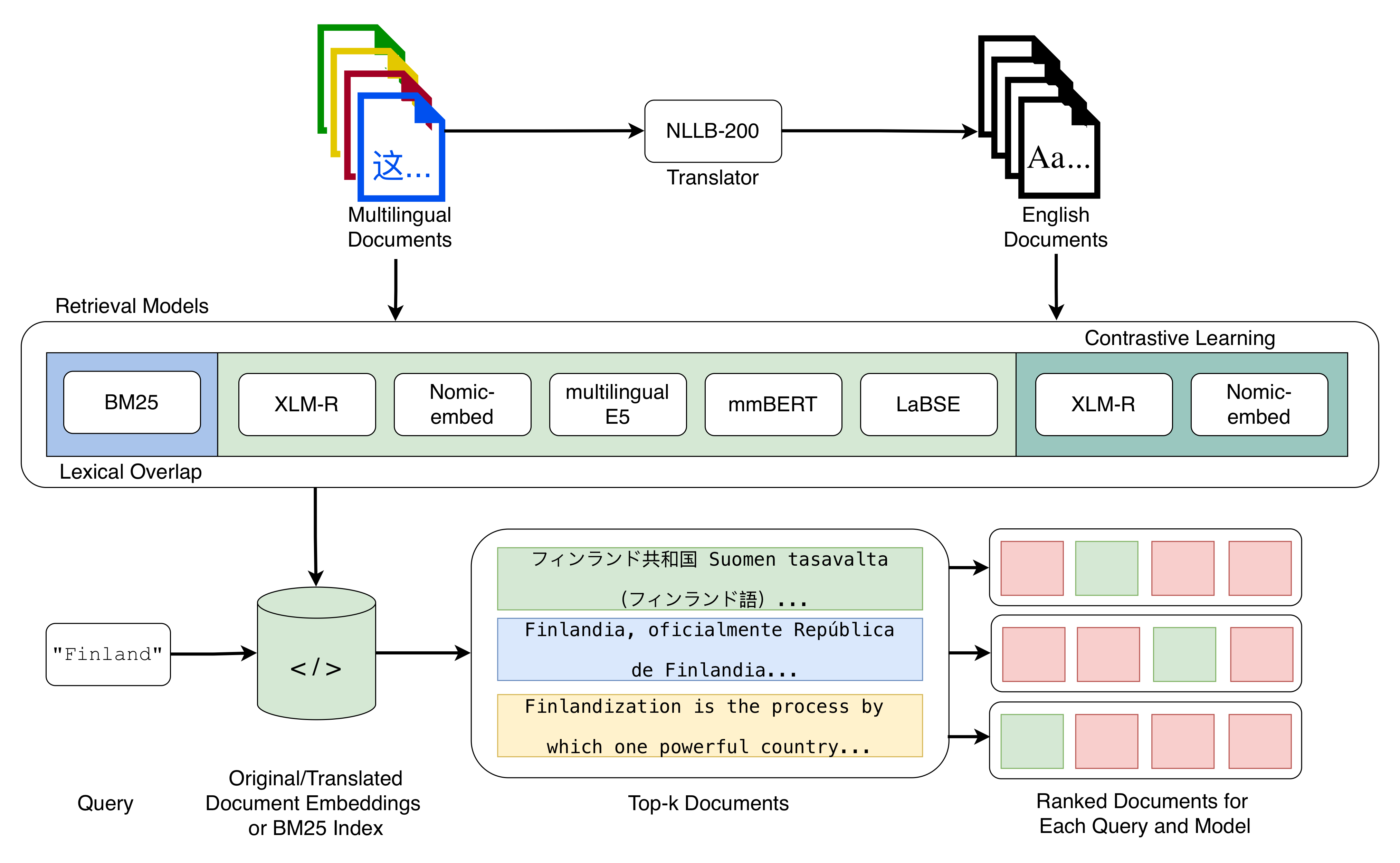}
    \caption{Overview of the CLIR ranking evaluation pipeline. }
    \label{fig:overview}
\end{figure}

\section{Introduction}

Cross-lingual information retrieval (CLIR) aims to retrieve documents written in one language in response to queries expressed in another \cite{sasaki-etal-2018-cross, goworek2025bridging}. As global information ecosystems expand, CLIR has become essential for equitable access to knowledge across linguistic boundaries \cite{oard1998crosslanguage, nie2010cross, litschko2022cross}. Nevertheless, it remains a challenging problem because of disparities in resources, script variation, morphological complexity, and uneven pretraining coverage across languages \cite{conneau-etal-2020-unsupervised, joshi2020state, xue2021mt5}. Methods that perform well in monolingual information retrieval often degrade in multilingual settings, as lexical mismatch and typological distance hinder effective comparison of queries and documents.

A common strategy in practical CLIR systems is to translate queries or documents into a pivot language, typically English, and then apply monolingual retrieval models \cite{sasaki-etal-2018-cross, lavrenko2002cross}. Although straightforward, this approach introduces noise from machine translation (MT), increases latency in large-scale systems, and does not guarantee that the resulting representations are aligned with the semantic space used for ranking \cite{fujii-ishikawa-2000-applying, zhou2021translational, xiao-guo-2013-semi}. Multilingual encoders provide a compelling alternative by embedding texts from different languages into a shared vector space, yet prior work has shown that their cross-lingual alignment quality varies widely across languages and domains \cite{pires2019multilingual, song-etal-2025-multilingual}.

Several linguistic and data-related factors compound the difficulty of CLIR. Lexical overlap is often non-existent across scripts; typological distance and morphological divergence reduce cross-lingual similarity; and pretraining corpora remain heavily skewed towards high-resource languages \cite{pires2019multilingual}. As a consequence, CLIR effectiveness can vary dramatically across language pairs, even for state-of-the-art multilingual encoders \cite{lin2023simple}.

Prior research has examined individual components of CLIR pipelines in isolation, such as translation quality \cite{sun2020clireval}, dense retrieval \cite{litschko2022cross}, or cross-encoder re-ranking \cite{litschko2022parameter}, and there is still limited understanding of how different interventions compare, especially when they are applied independently under controlled experimental conditions. In particular, it remains unclear when translation helps or harms retrieval, under which conditions contrastive alignment yields meaningful gains, whether re-ranking is effective for cross-script retrieval, and whether approximate nearest neighbour (ANN) \cite{Indyk1998ann} search offers efficiency benefits at the scale of multilingual document collections.

To address these gaps, this work evaluates several isolated interventions designed to improve encoder-based CLIR, focusing on four approaches: (i) embedding-based retrieval, comparing multilingual embedding models on native-language and English-translated corpora; (ii) contrastive alignment, applied at multiple linguistic granularities, namely, word, phrase, and alignment between queries and documents; and (iii) cross-encoder re-ranking with easy or hard negatives—where easy negatives are randomly sampled irrelevant documents, and hard negatives are documents that appear relevant to the query but are incorrect—used to train the cross-encoder to better discriminate fine-grained relevance; and a comparison of efficiency and effectiveness using ANN indexing. This experimental design makes it possible to determine when and why particular techniques yield improvements, and how these gains depend on linguistic factors, model alignment quality, and the underlying information retrieval architecture.

Figure \ref{fig:overview} illustrates the cross-lingual information retrieval evaluation pipeline used in this study. Multilingual documents are first translated into English using a machine translation system, for example NLLB-200 \cite{nllb_2022}. For each retrieval model, separate document embeddings and BM25 indices are then constructed over both the original and the translated corpora. During retrieval, the system selects the top 100 candidate documents for each query using either cosine similarity in the embedding space or BM25 lexical matching. Retrieval effectiveness is assessed over 1,000 or 500 queries per language pair using standard information retrieval metrics. This pipeline provides a unified framework for comparing lexical and semantic methods and for analysing how multilingual semantic representations influence retrieval performance across diverse languages, scripts and resource conditions. The main contributions are as follows:

\begin{itemize}
    \item \textbf{Evaluation of embedding-based CLIR methods.} This work presents a comparative evaluation of lexical, translation-based and  multilingual dense retrieval, and re-ranking techniques across three diverse datasets, examining their behaviour under varying resource levels and script conditions.
    
    \item \textbf{Semantic alignment in CLIR.} It analyses the role of semantic alignment \cite{litschko2022cross} in cross-lingual ranking by contrasting lexical overlap, translation and multilingual semantic modelling, and identifies the conditions under which contrastive alignment and cross-encoder re-ranking are most effective. 
    
    \item \textbf{Linguistic and efficiency factors.} It quantifies retrieval biases and typological correlations across language pairs and investigates efficiency mechanisms in encoder-based models. 
\end{itemize}

The remainder of this paper is structured as follows. Section \ref{sec:relatedwork} surveys the related work, beginning with lexical and translation-based approaches and outlining their limitations in cross-lingual settings, before introducing neural methods and semantic representations that currently underpin progress in the area. Section \ref{sec:methods} describes the experimental setup, including datasets, retrieval models, training configurations and evaluation protocols, so as to support clarity and reproducibility. Section \ref{sec:results} reports the empirical results and examines the effect of each intervention. It first highlights the limitations of lexical BM25 retrieval, including under document translation, and then contrasts these findings with the substantially stronger performance of multilingual language models, followed by an analysis of the contributions of contrastive learning and cross-encoder re-ranking to retrieval effectiveness. Section~\ref{sec:analysis} discusses the broader implications of the findings and presents an additional analysis of the linguistic factors that influence cross-lingual retrieval. Section \ref{sec:conclusion} concludes the paper by summarising the main insights and indicating directions for future work.

\section{Related Work}
\label{sec:relatedwork}

This section reviews prior research across three core dimensions of cross-lingual information retrieval (CLIR): (i) retrieval strategies including translation-based methods and modern sparse, dense, and hybrid architectures; (ii) multilingual semantic representations and alignment; and (iii) ranking strategies, including recent developments leveraging multilingual large language models (mLLMs).

\paragraph{Ranking Strategies.}
Early approaches to CLIR relied on lexical matching, typically by translating queries into the document language using bilingual dictionaries or statistical machine translation, followed by monolingual ranking models such as BM25 or query likelihood formulations \cite{oard1998crosslanguage,robertson1995okapi, mccarley1999should}. Document translation into a pivot language was also explored as an alternative, often yielding better disambiguation due to richer context \cite{mccarley1999should, kishida2003two}. 
However, translation-based pipelines are hindered by ambiguity, limited vocabulary coverage, and poor robustness to morphological variation, particularly for low-resource languages \cite{ballesteros_croft_1998, dadashkarimi2014probabilistic}. In addition, the translation step introduces latency and inconsistency in large-scale systems \cite{zhou2021translational, yao_2020}, making explicit query translation less feasible in operational settings. Consequently, recent research has shifted towards language-agnostic retrieval strategies that minimise dependence on machine translation by using shared embedding spaces.

Modern CLIR systems increasingly adopt neural retrieval models. Sparse neural retrievers such as SPLADE~\cite{formal2021splade} and SPLADE-v2~\cite{formal2021spladev2} enhance lexical matching via learned expansion, while dense retrieval approaches embed queries and documents into a shared vector space using multilingual encoders and rank via cosine similarity \cite{litschko2022cross, bonifacio_2022_mmarco, sun-duh-2020-clirmatrix}. Hybrid strategies combine sparse and dense signals to balance precision and semantic recall \cite{zhang2023modeling, khattab2020colbert}, with multilingual variants extending applicability across languages \cite{jha2024jina-colbert-v2}. To support scalability, particularly in dense retrieval, ANN search is employed to reduce inference overhead \cite{johnson2019billion, malkov2018efficient}, though its behaviour under cross-lingual conditions is less thoroughly explored \cite{wang2101comprehensive}. Multi-stage architectures are now prevalent, using sparse or dense retrieval for candidate generation, followed by more computationally intensive ranking models.  Cross-encoders jointly encode query and document for deeper token interaction and typically achieve high effectiveness \cite{nogueira2019passage, xiong2021approximate}, though at significant computational cost. Retrieval effectiveness, therefore, depends not only on the architecture but also on the quality of the underlying multilingual representation, which motivates the need for robust embedding alignment.

\paragraph{Multilingual Representations and Alignment.}
Multilingual encoders such as XLM-R~\cite{conneau-etal-2020-unsupervised} and mBERT~\cite{devlin_2019_bert} enabled shared embedding spaces for cross-lingual Natural Language Processing (NLP); however, alignment quality varies considerably across languages and typologies \cite{pires2019multilingual, song-etal-2025-multilingual}. Retrieval-oriented encoders improve alignment using ranking objectives and supervision from parallel or translation-based data, as in LaBSE \cite{feng-etal-2022-language}, LASER \cite{artetxe2018laser}, and multilingual-E5 \cite{wang2024multilingualE5}, achieving strong results on multilingual benchmarks such as MMTEB \cite{enevoldsen2025mmteb}. Contrastive learning is widely adopted in dense retrieval \cite{karpukhin-etal-2020-dense}, with supervision sourced from sentence-level alignment, translation pairs, or relevance labels from datasets such as mMARCO \cite{bonifacio_2022_mmarco}, CLIRMatrix \cite{sun-duh-2020-clirmatrix}, and MIRACL \cite{zhang_2023_miracl}. More recent work explores fine-grained alignment at token level using word sense or semantic similarity resources \cite{raganato2020xlwic, pilehvar2019wic, liu-etal-2021-am2ico}.

Although such models are increasingly used as the basis for the first-stage retrieval, multi-stage architectures rely on more expressive re-ranking mechanisms to refine document order, using their output as the candidate pool for more computationally intensive models.

\paragraph{Re-ranking with mLLMs.}
Recent work has applied mLLMs to CLIR re-ranking by exploiting their generative and reasoning capabilities. This includes query reformulation and expansion, relevance estimation via generative scoring or answer justification, and use as cross-lingual re-rankers that capture deeper semantic relations \cite{zuo-etal-2025-evaluating, huang-etal-2025-multilingual}. Beyond re-ranking, mLLMs have been used to reason directly over candidate passages, or to generate synthetic training data for retrievers \cite{muennighoff2023sgpt, min2025prompting, zhang2024promptingretrieval}.
mLLM-based re-ranking methods can be categorised as listwise, pairwise or pointwise. Listwise approaches prompt the model with multiple candidate passages simultaneously to estimate a ranked order \cite{ma2023lrl, sun-etal-2023-chatgpt}. Open-source variants such as RankVicuna and RankZephyr \cite{pradeep2023rankvicuna, pradeep2023rankzephyr}, and distillation-based methods like Rank-without-GPT \cite{zhang2025rankwithoutgpt}, demonstrate competitive performance on English benchmarks. Pairwise reranking compares candidate documents in pairs \cite{liu2025leveraging}. Pointwise methods independently assess query–document relevance or generate query expansions \cite{nogueira2020document, liang2023holistic, ma2023lrl}.
While these approaches improve cross-lingual semantic matching, they incur high computational cost, latency and reproducibility challenges. As a result, mLLMs are primarily used at the re-ranking or post-retrieval stage rather than for first-stage retrieval. Our work focuses on pointwise re-ranking architectures that are more computationally scalable while preserving multilingual effectiveness.

\section{Experimental Settings}
\label{sec:methods}

Cross-lingual retrieval is evaluated within a unified encoder-based pipeline. Four families of interventions are examined: (i) the effect of document translation on lexical and semantic retrieval; (ii) the impact of contrastive learning on multilingual alignment; (iii) the contribution of re-ranking over the first-stage retrieval; and (iv) the trade-off between retrieval effectiveness and efficiency.

\subsection{Datasets}
\label{sec:datasets}

Experiments are conducted on three established datasets: CLIRMatrix~\cite{sun-duh-2020-clirmatrix}, mMARCO~\cite{bonifacio_2022_mmarco} and the Large-Scale CLIR dataset~\cite{sasaki-etal-2018-cross}. CLIRMatrix covers all combinations of eight query and document languages, excluding same-language pairs, with documents from multilingual Wikipedia. mMARCO provides translations of MS~MARCO passages into 14 languages and supports all 14×14 language-pair combinations. The Large-Scale CLIR dataset consists of English queries paired with documents in 26 other languages. 

\begin{table*}[h]
\centering
\scriptsize
\setlength{\tabcolsep}{3pt}
\newcommand{\cmark}{\checkmark}

\begin{tabular}{l*{28}{c}}
\toprule
\textbf{Dataset} &
\textbf{AR} & \textbf{CA} & \textbf{CS} & \textbf{DE} & \textbf{EN} & \textbf{ES} &
\textbf{FI} & \textbf{FR} & \textbf{HI} & \textbf{ID} & \textbf{IT} & \textbf{JA} &
\textbf{KO} & \textbf{NL} & \textbf{NN} & \textbf{NO} & \textbf{PL} & \textbf{PT} &
\textbf{RO} & \textbf{RU} & \textbf{SV} & \textbf{SW} & \textbf{TL} & \textbf{TR} &
\textbf{UK} & \textbf{VI} & \textbf{ZH} \\
\midrule
CLIRMatrix &
\cmark &  &  & \cmark & \cmark & \cmark &
 & \cmark &  &  &  & \cmark &
  &  &  &  &  &  &  & \cmark &  &  &  &  &  &  & \cmark \\
mMARCO &
\cmark &  &  & \cmark & \cmark & \cmark &
 & \cmark & \cmark & \cmark & \cmark & \cmark &
  & \cmark &  &  &  & \cmark &  & \cmark &  &  &  &  &  & \cmark & \cmark \\
Large-Scale &
\cmark & \cmark & \cmark & \cmark & \cmark & \cmark &
\cmark & \cmark &  &  & \cmark & \cmark &
\cmark & \cmark & \cmark & \cmark & \cmark & \cmark & \cmark & \cmark &
\cmark & \cmark & \cmark & \cmark & \cmark & \cmark & \cmark \\
\bottomrule
\end{tabular}
\caption{Language coverage of the three datasets. Columns list 2-letter ISO codes. A tick (\cmark) indicates inclusion. mMARCO and CLIRMatrix are cross-lingual (CLIRMatrix excludes same-language pairs), while Large-Scale provides only English queries, paired with documents in the listed languages.}
\label{tab:dataset-language-coverage}
\end{table*}

\newcommand{\cmark}{\checkmark}

Table \ref{tab:dataset-language-coverage} summarises the language coverage of the three datasets, with both queries and documents available in all languages with $\cmark$ in Table~\ref{tab:dataset-language-coverage} for CLIRMatrix and mMARCO, but only English queries, matched with documents in all the specified languages for Large-Scale. 

This work uses the following ISO 639-1 language codes in the experiments: AR (Arabic), CA (Catalan), CS (Czech), DE (German), EN (English), 
ES (Spanish), FI (Finnish), FR (French), HI (Hindi), ID (Indonesian), 
IT (Italian), JA (Japanese), KO (Korean), NL (Dutch), NN (Norwegian Nynorsk), 
NO (Norwegian Bokmål), PL (Polish), PT (Portuguese), RO (Romanian), 
RU (Russian), SV (Swedish), SW (Swahili), TL (Tagalog), TR (Turkish), 
UK (Ukrainian), VI (Vietnamese), and ZH (Chinese).

To ensure comparability, all datasets are evaluated with a top-100 retrieval depth. In CLIRMatrix, 1{,}000 queries are sampled for each of the 56 language pairs, yielding 56{,}000 query instances and 9{,}055 unique documents (1{,}080–1{,}205 per language). In mMARCO, translations are deduplicated across languages and the resulting 7{,}433 unique documents are evenly redistributed across all 14 languages, assigning 530–531 queries per pair for a total of 97{,}720 query–document pairs. In the Large-Scale dataset, each English$\rightarrow$X language pair includes 1{,}000 queries, giving 26{,}000 query pairs overall. For each dataset, top-100 document retrieval for each query is performed over the full multilingual document pool, and all documents are truncated to 512 tokens.

\subsection{Retrieval Pipeline}
\label{sec:models}

This section introduces the pretrained multilingual encoders used as retrieval models in the cross-lingual experiments. 

\paragraph{Retrieval Models.} Five multilingual encoders are considered, spanning diverse pretraining and alignment paradigms and covering around 100 languages each. XLM-R~\cite{conneau-etal-2020-unsupervised} serves as a strong masked language modelling baseline, trained on 100 languages with approximately 550 million parameters. LaBSE~\cite{feng-etal-2022-language} is a dual encoder trained with a translation-ranking objective, supporting 109 languages with 471 million parameters. Multilingual-E5~\cite{wang2024multilingualE5} is an information retrieval-oriented extension of the E5 model with multilingual instruction tuning, covering more than 100 languages and comprising 560 million parameters. Nomic-embed-text-v2-moe~\cite{nussbaum2025training} (``Nomic'') is a sparse mixture-of-experts model with coverage of roughly 100 languages and an effective parameter budget of about 475 million. Finally, mmBERT~\cite{marone2025mmbertmodernmultilingualencoder} is a recent multilingual BERT variant with improved cross-lingual alignment and vocabulary coverage, supporting over 100 languages with approximately 307 million parameters.

\paragraph{Document Translation.} To improve comparability across languages, all documents in CLIRMatrix and Large-Scale are translated into English using NLLB-200 \cite{nllb_2022}. For mMARCO, which already contains parallel documents in all languages, the corresponding English passages are used for the translated condition.
The translation model is selected based on a pilot evaluation of four MT systems on a sample of 100 documents per language. Two criteria are used: COMET~\cite{rei-etal-2020-comet}, which estimates translation adequacy, and the perplexity of LLaMA-3.1-8B~\cite{llama3_2024}, which reflects translation fluency. The systems considered are DeepL \cite{deepl2024}, a strong commercial MT service; NLLB-200, Meta’s open multilingual model covering 200 languages; googletrans \cite{googletrans2024}, a lightweight Python interface to Google Translate; and LibreTranslate \cite{libretranslate2024}, an open-source MT service that can be run locally. NLLB-200 is chosen as it consistently ranks near the top (typically second best across languages) on both adequacy and fluency. DeepL achieves the best scores overall, but rate limits and cost make it unsuitable for large-scale experiments, so it is treated as an approximate upper bound rather than a deployable option. 

\paragraph{First-stage Ranking.} Ranking is performed using a bi-encoder architecture. For each encoder, document embeddings are precomputed offline. At inference time, query embeddings are obtained and documents are ranked according to the cosine similarity between query and document embeddings, retaining the indices of the top 100 documents for each query. As a lexical baseline, BM25~\cite{robertson2009probabilistic} is included, with the document index constructed in advance and queried at retrieval time. This stage requires no additional training and serves as a control condition for assessing model- and data-level interventions. All encoders use mean pooling over the final hidden layer, and the resulting embeddings are L2-normalised. No adapters or projection layers are added in the base experiments, and retrieval relies on cosine similarity throughout.

\paragraph{Contrastive Learning Settings.} Fine-tuning is applied only to the weakest (XLM-R) and strongest (Nomic) pretrained encoders in order to assess whether the effectiveness of contrastive interventions depends on baseline multilingual alignment. Contrastive learning is carried out using text pairs at three levels of granularity. At the word level, training uses multilingual word sense disambiguation datasets (XL-WiC~\cite{raganato2020xlwic}, Am$^{2}$iCO~\cite{liu-etal-2021-am2ico}, SenWiCh~\cite{goworek2025senwich} restricted to languages that overlap with those in the retrieval datasets. At the phrase level, up to 1,000 Tatoeba~\cite{tiedemann-2020-tatoeba} parallel sentences are used for each language pair present in the retrieval benchmarks. At the query–document level, fine-tuning is performed on relevance-annotated pairs from CLIRMatrix, mMARCO and the Large-Scale CLIR dataset. 

\paragraph{Cross-encoder Re-ranking Settings.} XLM-R and Nomic are further fine-tuned as cross-encoders that jointly encode each query–document pair and output a relevance score via a classification head. Re-ranking operates over the top-100 BM25 candidates; if the relevant document is absent from this candidate set, it replaces the item at rank 100. The classifier is trained with binary relevance labels, and two negative sampling strategies are compared: \textit{easy negatives}, obtained by randomly sampling non-relevant documents, and \textit{hard negatives}, obtained from non-relevant documents retrieved by the first-stage ranker and therefore harder to distinguish from true positives (available for CLIRMatrix and mMARCO). 

Further details of the experimental settings and implementation are provided in Appendix~\ref{app:models_performance_translated}-~\ref{app:ling-sim}.

\subsection{Evaluation}

For each query, a ranked list of candidate documents is obtained using cosine similarity between L2-normalised bi-encoder embeddings, which serves as the query–document similarity score for all neural retrieval models. For BM25, document ranking is performed directly over the inverted index using the query tokens. In both cases, the top-100 documents are retained as the retrieval output for evaluation.

Retrieval effectiveness is measured over 1{,}000 (or 500) queries per language pair using standard IR metrics. The primary metric is Recall@100, defined as the proportion of queries for which the relevant document appears in the top-100 results. For re-ranking experiments, Recall@10 and nDCG@100 (normalized Discounted Cumulative Gain) \cite{jarvelin2002cumulated} are additionally reported, where nDCG captures the quality of the ranking by assigning higher weight to relevant documents appearing near the top of the list and normalising scores across queries. Metrics are averaged over all language pairs within each dataset, and four representative pairs (en–es, en–zh, ja–en, ar–ru) are highlighted to illustrate behaviour across high- and low-resource, same- and cross-script settings. Finally, post-hoc analyses examine (i) the relationship between ANN latency and retrieval accuracy; (ii) document-language retrieval bias; and (iii) correlations between retrieval quality and linguistic similarity.

A summary of the experiments, covering the models, training regimes, levels of comparison and types of retrieval can be seen in Table~\ref{tab:summary}.

\begin{table}[!h]
\centering
\begin{tabular}{llll}
\toprule
\textbf{Encoder(s)} & \textbf{Training} & \textbf{Level} & \textbf{Retrieval Type}\\
\midrule
BM25 & -- & Original/Translated & Lexical \\
5 pretrained multilingual encoders & -- & Original/Translated & Cosine \\
Nomic, XLM-R & Contrastive & Word/Phrase/QD & Cosine  \\
Nomic, XLM-R & Easy/Hard Negatives & QD & Cross-Encoder  \\
5 pretrained multilingual encoders & - & exact/ANN & Cosine \\
\bottomrule
\end{tabular}
\caption{Summary of experimental settings across all intervention types. (QD = Query-Document)}
\label{tab:summary}
\end{table}

ANN search is implemented using Hierarchical Navigable Small Worlds (HNSW) \cite{malkov2018efficient} indices in order to measure potential speed-ups over exact cosine-similarity search. All ANN experiments compare latency and Recall@100 against exhaustive nearest-neighbour retrieval on the same embeddings. Index construction details and parameter settings are reported in Appendix~\ref{app:ann}.

\section{Results}
\label{sec:results}

This section first selects the MT model used for document translation, then evaluates lexical BM25 retrieval with and without translation. It then examines the retrieval performance of pretrained multilingual encoders, before analysing the effects of contrastive fine-tuning, cross-encoder re-ranking and ANN search on effectiveness and efficiency.

\subsection{Machine Translation Model}

Candidate MT systems are evaluated to select the model used for document translation in the multilingual experiments. Translation quality is assessed on a subset of 100 documents per language from CLIRMatrix and Large-Scale dataset, after translating documents into English. Two criteria are used: COMET~\cite{rei-etal-2020-comet}, which estimates translation adequacy, and the perplexity of LLaMA-3.1-8B~\cite{llama3_2024}, which reflects the fluency of the translated text. For the mMARCO dataset, the already available English documents are used in the translated condition, as the collection contains parallel queries and documents across all languages.

\begin{figure}[h]
    \centering
    \includegraphics[width=0.9\linewidth]{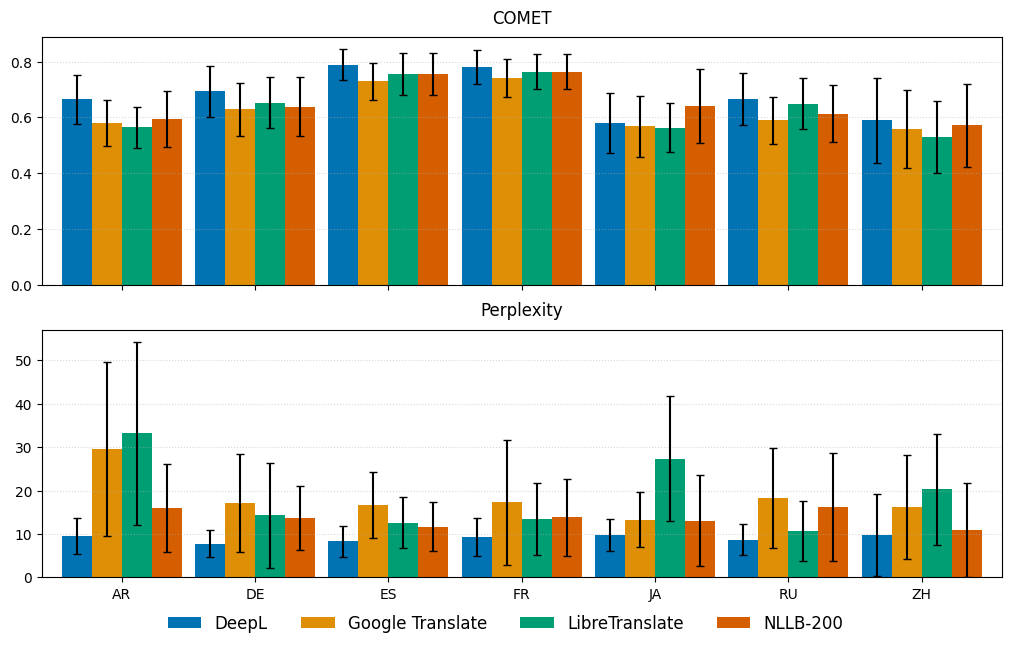}
    \caption{Translation quality of different models evaluated with COMET and Perplexity, by translating non-English documents from the CLIRMatrix dataset \cite{sun-duh-2020-clirmatrix}, sampling the same 100 documents from each language. }
    \label{fig:translation_quality_by_model_language}
\end{figure}

Figure ~\ref{fig:translation_quality_by_model_language} shows that DeepL consistently achieves the highest COMET scores and lowest perplexity. Because its free tier is limited to roughly 500k characters per month, the available quota is used twice to complete the evaluation. Among free and open-source systems, NLLB-200 emerges as the strongest option, outperforming googletrans and LibreTranslate on both adequacy and fluency. Googletrans is also less reliable in this setting, failing to translate two Arabic and one French document that contain extensive English code-switching, which triggers translation failures. On this basis, NLLB-200 is selected to translate the document corpora for subsequent experiments.

Across MT systems, Spanish and French consistently obtain low perplexity and high COMET scores, whereas Arabic, Japanese and Chinese show substantially lower COMET scores and highly variable perplexity. These differences reflect typological and script-related biases in both translation models and evaluation metrics. Script divergence in particular emerges as a persistent source of difficulty, with non-Latin scripts yielding less stable and generally poorer translations. This pattern aligns with broader observations that cross-script transfer remains a major bottleneck for multilingual NLP and contributes to uneven downstream cross-lingual retrieval performance.

\subsection{Cross-Lingual Information Retrieval}

After selecting the translation model for corpus normalisation, retrieval performance is examined under both translated and non-translated conditions. As a first step, BM25 is evaluated to characterise the limitations of purely lexical matching before moving to semantic encoders. Comparisons of BM25 with and without document translation confirm that lexical matching is largely ineffective for cross-lingual retrieval, particularly for non-English and cross-script language pairs.

\begin{figure*}[h!]
\centering
\begin{subfigure}{0.49\linewidth}
\centering
\includegraphics[width=\linewidth]{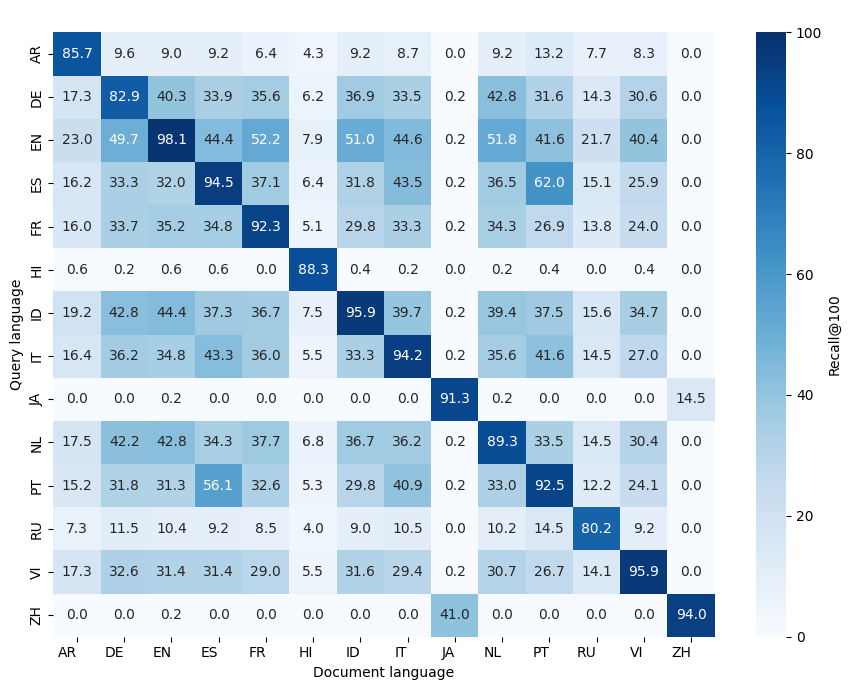}
\caption{}
\label{fig:bm25_mmarco}
\end{subfigure}
\hfill
\begin{subfigure}{0.49\linewidth}
\centering
\includegraphics[width=\linewidth]{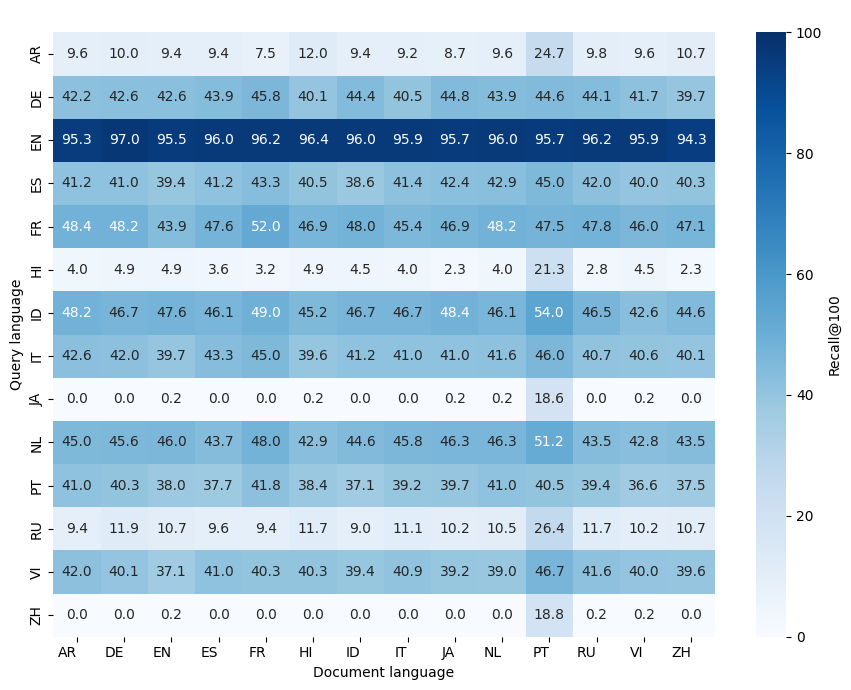}
\caption{}
\label{fig:bm25_mmarco_translated}
\end{subfigure}
\caption{Comparison of lexical BM25 retrieval (Recall@100) on query–document language
pairs for cross-lingual retrieval (a) without and (b) with document translation on the mMARCO dataset \cite{bonifacio_2022_mmarco}.}
\label{fig:bm25_comparison}
\end{figure*}


In operational settings, where queries may arrive in unknown languages or MT coverage is limited, reliance on BM25 without translation reduces effectiveness to near zero for most non-English pairs. Figure~ \ref{fig:bm25_comparison} illustrates this limitation. Without translation (Figure~ \ref{fig:bm25_mmarco}), BM25 retrieves almost no relevant documents for non-English pairs, as expected for methods that depend on shared lexical forms. Translating all documents into English improves performance only for English queries (Figure~ \ref{fig:bm25_mmarco_translated}) by restoring lexical overlap between English queries and English documents, but it does not benefit queries in other languages and remains ineffective in cross-script settings. These findings motivate the move from lexical matching to semantic embedding-based ranking.

\begin{figure*}[h!]
\centering
\includegraphics[width=\linewidth]{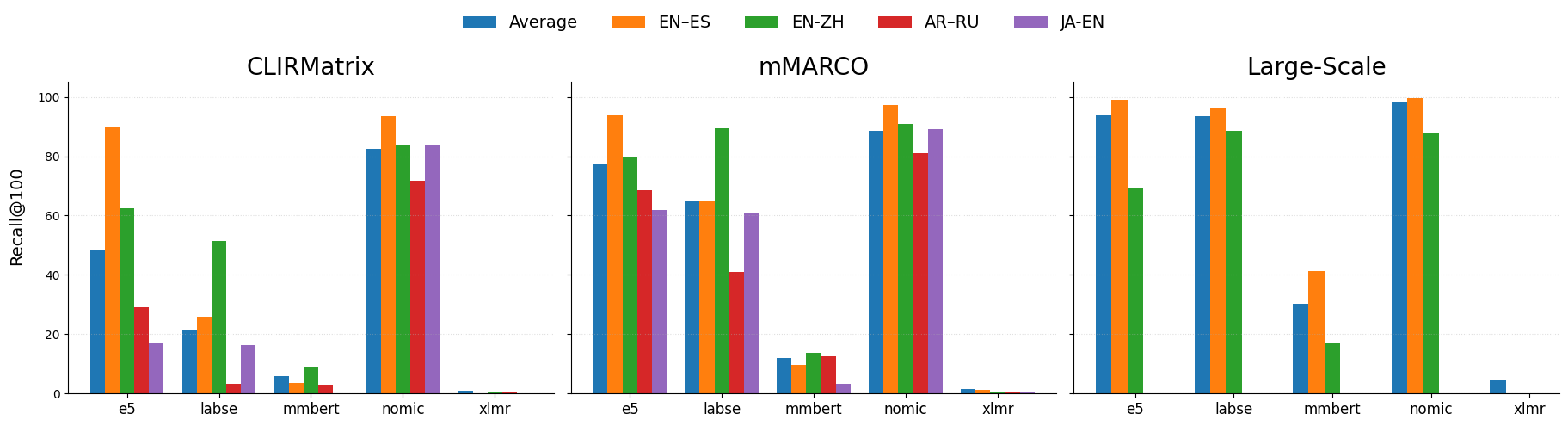}
\caption{Performance (Recall@100) of five pretrained multilingual encoders using cosine similarity between query and document embeddings without translation. Results are averaged across all language pairs and shown for selected pairs.}
\label{fig:models_performance}
\end{figure*}

\paragraph{Semantic Embedding-Based Ranking.} Given the weakness of lexical retrieval in multilingual settings, attention turns to the performance of pretrained multilingual encoders. Five encoders are compared across datasets and representative language pairs. This comparison shows that dense models dramatically outperform BM25 across all benchmarks. As displayed in Figure~ \ref{fig:models_performance}, Nomic consistently achieves the strongest performance, while XLM-R is the weakest. These two encoders therefore serve as representative extremes for analysing the effects of contrastive learning at different levels of granularity.

Performance patterns across language pairs reveal persistent difficulties for distant or low-resource combinations. English–Spanish is generally the easiest direction for most models (with the exception of LaBSE), whereas Arabic–Russian is consistently the most challenging, highlighting limitations in bridging typologically distant languages and cross-script retrieval. Evaluating the same models on a fully translated corpus (Appendix~ \ref{app:models_performance_translated}) yields only minor changes: small improvements for some encoders and small declines for others.

\newcommand{\cc}[1]{\gradientcelld{#1}{-100}{0}{100}{red}{white}{green}{70}} 

\begin{table}[h] 
\centering 
\small 
\begin{tabular}{lcc} 
\toprule & \textbf{Original Documents} & \textbf{Translated Documents} \\ 
\midrule \multicolumn{3}{l}{\textbf{CLIRMatrix}} \\ \quad multilingual-E5 & \cc{48.3} & \cc{19.5} \\ \quad LaBSE & \cc{21.4} & \cc{19.0} \\ \quad mmBERT & \cc{6.0} & \cc{4.2} \\ \quad Nomic & \cc{82.6} & \cc{84.7} \\ \quad XLM-R & \cc{1.0} & \cc{0.9} \\ \midrule \multicolumn{3}{l}{\textbf{mMARCO}} \\ \quad multilingual-E5 & \cc{77.5} & \cc{94.7} \\ \quad LaBSE & \cc{65.2} & \cc{69.6} \\ \quad mmBERT & \cc{12.1} & \cc{10.2} \\ \quad Nomic & \cc{88.6} & \cc{92.2} \\ \quad XLM-R & \cc{1.6} & \cc{1.9} \\ \midrule \multicolumn{3}{l}{\textbf{Large-Scale}} \\ \quad multilingual-E5 & \cc{93.8} & \cc{93.1} \\ \quad LaBSE & \cc{93.5} & \cc{86.9} \\ \quad mmBERT & \cc{30.3} & \cc{34.9} \\ \quad Nomic & \cc{98.3} & \cc{93.9} \\ \quad XLM-R & \cc{4.3} & \cc{6.9} \\ 
\bottomrule 
\end{tabular} 
\caption{Impact of document translation on embedding-based retrieval. Average Recall@100 scores across all language pairs for each dataset when using document embeddings in their original language versus translated into English.} 
\label{tab:avg_impact_of_translation} 
\end{table}

Document translation has a limited effect on embedding-based retrieval, with most models showing small or inconsistent changes in Recall@100 across datasets. In contrast to BM25, where translation is essential to restore lexical overlap, dense encoders already capture cross-lingual semantic similarity, making their performance largely invariant to whether documents are embedded in their original language or in English. Overall, these results indicate that multilingual encoders rely primarily on semantic rather than surface-level cues, and that document translation offers little additional benefit in dense retrieval pipelines. This motivates a closer examination of whether targeted contrastive alignment can further improve retrieval effectiveness.

\paragraph{Contrastive Model Fine-Tuning.}

Table~ \ref{tab:contrastive_impact} shows that contrastive fine-tuning affects XLM-R and Nomic in markedly different ways, depending on the level of supervision. For XLM-R, phrase-level alignment brings substantial gains, suggesting that its pretrained representations do not fully capture cross-lingual semantic relations and remain sensitive to language-specific noise. Query–document supervision yields by far the strongest improvements, increasing Recall@100 from 1.0\% 4.3\% and 1.6\% in the baseline to 21.2\%, 96.0\% and 77.1\% on CLIRMatrix, Large-Scale and mMARCO respectively, which indicates that retrieval-specific objectives are essential for general-purpose multilingual encoders. In contrast, word-level alignment has no meaningful effect, likely because query terms are rarely ambiguous in context and isolated word embeddings provide insufficient information for disambiguation.

For Nomic, which is already trained with cross-lingual alignment and retrieval objectives, additional contrastive fine-tuning has limited impact. Performance remains above 80\% across most strategies, except with word-level alignment, and only query–document supervision on CLIRMatrix yields a modest improvement of around \textasciitilde+4\%. This suggests that Nomic is already well optimised for cross-lingual retrieval and that further supervision produces diminishing returns rather than systematic gains.

\begin{table}[h!]
\centering
\begin{tabular}{llrrrr}
\toprule
\textbf{Model} & \textbf{Dataset} & \textbf{None} & \textbf{Word} & \textbf{Phrase} & \textbf{Query--Doc} \\
\midrule
\multirow{3}{*}{XLM-R} 
    & CLIRMatrix   & \cc{1.0}  & \cc{1.1}  & \cc{13.7} & \cc{21.2} \\
    & Large-Scale  & \cc{4.3}  & \cc{0.4}  & \cc{64.8} & \cc{96.0} \\
    & mMARCO       & \cc{1.6}  & \cc{1.4}  & \cc{68.0} & \cc{77.1} \\
\midrule
\multirow{3}{*}{Nomic}
    & CLIRMatrix   & \cc{82.6} & \cc{31.1} & \cc{79.9} & \cc{88.3} \\
    & Large-Scale  & \cc{98.3} & \cc{98.3} & \cc{98.3} & \cc{97.2} \\
    & mMARCO       & \cc{88.6} & \cc{88.6} & \cc{88.6} & \cc{84.4} \\
\bottomrule
\end{tabular}
\caption{Effect of contrastive fine-tuning (word, phrase, query–document level) on cosine-similarity retrieval (Recall@100\,\%) using XLM-R and Nomic, compared against non–fine-tuned baselines (None) across datasets, averaged across all language pairs.}
\label{tab:contrastive_impact}
\end{table}

\begin{figure*}[h!]
    \centering
    \includegraphics[width=\linewidth]{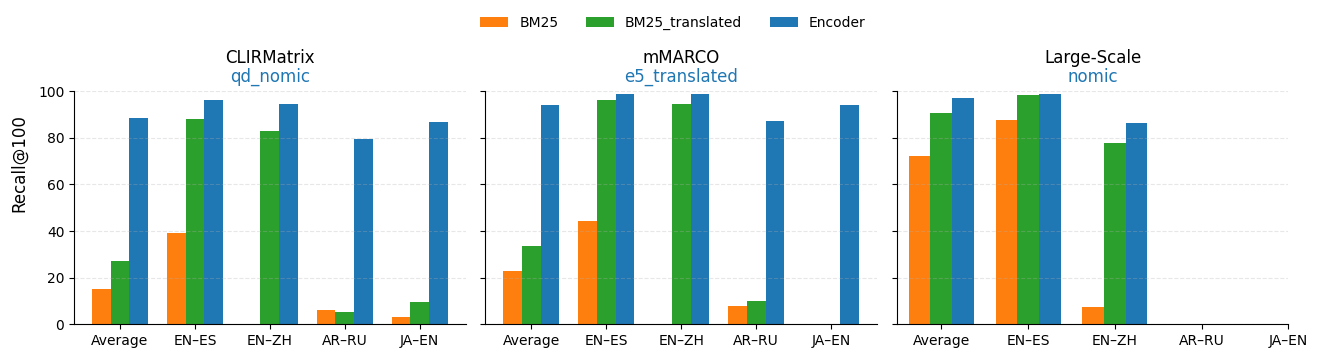}
    \caption{Performance of the best-performing embedding-based model on each dataset and BM25, with and without document translation, on average, and across selected query–document language pairs.}
    \label{fig:Best_over_bm25}
\end{figure*}

To contextualise the benefits of semantic encoders, the best-performing dense model on each dataset is contrasted with BM25. Figure~ \ref{fig:Best_over_bm25} compares the strongest embedding-based configuration with BM25 applied to both original and English-translated corpora. Improvements over lexical retrieval are substantial across all datasets and are especially pronounced for low-resource and script-divergent language pairs, where BM25 fails almost entirely because of the lack of shared vocabulary. Dense retrievers also outperform BM25 when both languages are relatively well resourced, such as English–Spanish and English–Chinese, demonstrating that semantic representations provide benefits beyond vocabulary mismatch. These gains persist regardless of whether BM25 is applied to original or translated documents.

\begin{table}[h]
\centering
\small
\begin{tabular}{lccccc}
\toprule
Dataset     & Average & EN-ES   & EN-ZH   & AR-RU   & JA-EN   \\
\midrule
\multirow{2}{*}{CLIRMatrix}
  & QD\_N   & QD\_N   & QD\_N   & QD\_N   & QD\_N\_T \\
  & 88.3   & 96.2   & 94.6   & 79.6   & 87.1    \\
\midrule
\multirow{2}{*}{mMARCO}
  & E5\_T   & N\_T    & N\_T    & E5\_T   & E5\_T    \\
  & 94.7   & 99.6   & 99.6   & 88.5   & 95.3    \\
\midrule
\multirow{2}{*}{Large-Scale}
  & N       & N       & N\_T    & ---     & ---      \\
  & 98.3   & 99.7   & 91.3   &         &          \\
\bottomrule
\end{tabular}
\caption{Best performing model on each dataset, and in the selected language pairs, and the Recall@100 score.
QD = query-document contrastive learning,\;
T = on translated documents,\;
N = Nomic,\;
E5 = multilingual-E5..}
\label{tab:best_models_table}
\end{table}

Table~ \ref{tab:best_models_table} summarises which model achieves the highest Recall@100 for each dataset, both on average and for selected language pairs. On CLIRMatrix, query–document contrastive learning substantially improves the already strong Nomic encoder, yielding the best results across all evaluated language pairs. The notably high performance observed for the Japanese–English pair in the translated condition does not follow from improved translation quality, since these documents are originally in English and remain untranslated. Instead, the improvement stems from reducing the influence of document embeddings in other languages that, in the non-translated setting, occasionally rank above the correct English documents. For mMARCO, both translation and strong cross-lingual alignment contribute to improved retrieval quality. This pattern is expected: the ``multilingual” mMARCO variants are created by translating the original English MS MARCO passages, so retrieval over the original English corpus recovers the highest-quality texts, while Nomic and multilingual-E5 already provide strong cross-lingual alignment (as reflected in multilingual-E5’s state-of-the-art performance on MMTEB~ \cite{enevoldsen2025mmteb}), leaving little room for further gains through fine-tuning. Finally, on the Large-Scale dataset, Nomic without any additional interventions achieves the best overall performance. Given that Nomic is explicitly optimised for cross-lingual retrieval and already attains very high effectiveness, further improvements are naturally hard to obtain.


\newcommand{\defbm}[4]{%
  \expandafter\def\csname #1_low\endcsname{#2}%
  \expandafter\def\csname #1_middle\endcsname{#3}%
  \expandafter\def\csname #1_high\endcsname{#4}%
}

\newcommand{\g}[2][]{%
  \gradientcelld{#2}%
    {\csname #1_low\endcsname}%
    {\csname #1_middle\endcsname}%
    {\csname #1_high\endcsname}%
    {red}{white}{green}{70}%
}


\defbm{CLIRMatrix_R10_avg}{-39.6}{10.4}{60.4}
\defbm{CLIRMatrix_R10_enes}{-18.6}{31.4}{81.4}
\defbm{CLIRMatrix_R10_enzh}{-5}{0}{5}
\defbm{CLIRMatrix_R10_arru}{-48.3}{1.7}{51.7}
\defbm{CLIRMatrix_R10_jaen}{-46.9}{3.1}{53.1}

\defbm{CLIRMatrix_NDCG_avg}{-43.8}{6.2}{56.2}
\defbm{CLIRMatrix_NDCG_enes}{-33.1}{16.9}{66.9}
\defbm{CLIRMatrix_NDCG_enzh}{-5}{0}{5}
\defbm{CLIRMatrix_NDCG_arru}{-48.4}{1.6}{51.6}
\defbm{CLIRMatrix_NDCG_jaen}{-48.5}{1.5}{51.5}

\defbm{mMARCO_R10_avg}{-32.7}{17.3}{67.3}
\defbm{mMARCO_R10_enes}{-20.2}{29.8}{79.8}
\defbm{mMARCO_R10_enzh}{-5}{0}{5}
\defbm{mMARCO_R10_arru}{-44.7}{5.3}{55.3}
\defbm{mMARCO_R10_jaen}{-49.8}{0.2}{50.2}

\defbm{mMARCO_NDCG_avg}{-35.1}{14.9}{64.9}
\defbm{mMARCO_NDCG_enes}{-23.5}{26.5}{76.5}
\defbm{mMARCO_NDCG_enzh}{-5}{0}{5}
\defbm{mMARCO_NDCG_arru}{-45.7}{4.3}{54.3}
\defbm{mMARCO_NDCG_jaen}{-49.8}{0.2}{50.2}

\defbm{LargeScale_R10_avg}{13.5}{63.5}{100}
\defbm{LargeScale_R10_enes}{26.3}{76.3}{100}
\defbm{LargeScale_R10_enzh}{-42.9}{7.1}{57.1}

\defbm{LargeScale_NDCG_avg}{7.9}{57.9}{100}
\defbm{LargeScale_NDCG_enes}{19.7}{69.7}{100}
\defbm{LargeScale_NDCG_enzh}{-43.5}{6.5}{56.5}


\begin{table}[]
\centering
\resizebox{\textwidth}{!}{%
\begin{tabular}{lllrrrrrrrrrr}
\toprule
\multirow{2}{*}{Dataset} & \multirow{2}{*}{Model} & \multirow{2}{*}{Negatives}
  & \multicolumn{5}{c}{Recall@10}
  & \multicolumn{5}{c}{nDCG@100} \\
\cmidrule(lr){4-8}\cmidrule(lr){9-13}
 &  &  & Avg & EN-ES & EN-ZH & AR-RU & JA-EN
   & Avg & EN-ES & EN-ZH & AR-RU & JA-EN \\
\midrule
\multirow{5}{*}{CLIRMatrix}
 & BM25  & --    &
   \g[CLIRMatrix_R10_avg]{10.4} &
   \g[CLIRMatrix_R10_enes]{31.4} &
   \g[CLIRMatrix_R10_enzh]{0.0} &
   \g[CLIRMatrix_R10_arru]{1.7} &
   \g[CLIRMatrix_R10_jaen]{03.1} &
   \g[CLIRMatrix_NDCG_avg]{6.2} &
   \g[CLIRMatrix_NDCG_enes]{16.9} &
   \g[CLIRMatrix_NDCG_enzh]{0.0} &
   \g[CLIRMatrix_NDCG_arru]{01.6} &
   \g[CLIRMatrix_NDCG_jaen]{01.5} \\
 & Nomic & easy  &
   \g[CLIRMatrix_R10_avg]{11.8} &
   \g[CLIRMatrix_R10_enes]{10.0} &
   \g[CLIRMatrix_R10_enzh]{60.0} &
   \g[CLIRMatrix_R10_arru]{0.0} &
   \g[CLIRMatrix_R10_jaen]{10.0} &
   \g[CLIRMatrix_NDCG_avg]{21.5} &
   \g[CLIRMatrix_NDCG_enes]{23.0} &
   \g[CLIRMatrix_NDCG_enzh]{41.1} &
   \g[CLIRMatrix_NDCG_arru]{17.0} &
   \g[CLIRMatrix_NDCG_jaen]{23.2} \\
 & Nomic & hard  &
   \g[CLIRMatrix_R10_avg]{33.7} &
   \g[CLIRMatrix_R10_enes]{30.0} &
   \g[CLIRMatrix_R10_enzh]{30.0} &
   \g[CLIRMatrix_R10_arru]{30.0} &
   \g[CLIRMatrix_R10_jaen]{50.0} &
   \g[CLIRMatrix_NDCG_avg]{29.4} &
   \g[CLIRMatrix_NDCG_enes]{28.6} &
   \g[CLIRMatrix_NDCG_enzh]{27.8} &
   \g[CLIRMatrix_NDCG_arru]{26.2} &
   \g[CLIRMatrix_NDCG_jaen]{31.8} \\
 & XLM-R & easy  &
   \g[CLIRMatrix_R10_avg]{35.0} &
   \g[CLIRMatrix_R10_enes]{50.0} &
   \g[CLIRMatrix_R10_enzh]{30.0} &
   \g[CLIRMatrix_R10_arru]{30.0} &
   \g[CLIRMatrix_R10_jaen]{50.0} &
   \g[CLIRMatrix_NDCG_avg]{33.3} &
   \g[CLIRMatrix_NDCG_enes]{33.9} &
   \g[CLIRMatrix_NDCG_enzh]{27.2} &
   \g[CLIRMatrix_NDCG_arru]{26.8} &
   \g[CLIRMatrix_NDCG_jaen]{49.0} \\
 & XLM-R & hard  &
   \g[CLIRMatrix_R10_avg]{30.2} &
   \g[CLIRMatrix_R10_enes]{10.0} &
   \g[CLIRMatrix_R10_enzh]{20.0} &
   \g[CLIRMatrix_R10_arru]{60.0} &
   \g[CLIRMatrix_R10_jaen]{20.0} &
   \g[CLIRMatrix_NDCG_avg]{26.3} &
   \g[CLIRMatrix_NDCG_enes]{23.0} &
   \g[CLIRMatrix_NDCG_enzh]{26.2} &
   \g[CLIRMatrix_NDCG_arru]{33.3} &
   \g[CLIRMatrix_NDCG_jaen]{30.3} \\
\midrule
\multirow{5}{*}{mMARCO}
 & BM25  & --    &
   \g[mMARCO_R10_avg]{17.3} &
   \g[mMARCO_R10_enes]{29.8} &
   \g[mMARCO_R10_enzh]{0.0} &
   \g[mMARCO_R10_arru]{05.3} &
   \g[mMARCO_R10_jaen]{0.2} &
   \g[mMARCO_NDCG_avg]{14.9} &
   \g[mMARCO_NDCG_enes]{26.5} &
   \g[mMARCO_NDCG_enzh]{0.0} &
   \g[mMARCO_NDCG_arru]{04.3} &
   \g[mMARCO_NDCG_jaen]{0.2} \\
 & Nomic & easy  &
   \g[mMARCO_R10_avg]{13.6} &
   \g[mMARCO_R10_enes]{20.0} &
   \g[mMARCO_R10_enzh]{0.0} &
   \g[mMARCO_R10_arru]{06.0} &
   \g[mMARCO_R10_jaen]{0.0} &
   \g[mMARCO_NDCG_avg]{25.1} &
   \g[mMARCO_NDCG_enes]{30.1} &
   \g[mMARCO_NDCG_enzh]{18.0} &
   \g[mMARCO_NDCG_arru]{19.8} &
   \g[mMARCO_NDCG_jaen]{15.4} \\
 & Nomic & hard  &
   \g[mMARCO_R10_avg]{08.7} &
   \g[mMARCO_R10_enes]{18.0} &
   \g[mMARCO_R10_enzh]{04.0} &
   \g[mMARCO_R10_arru]{03.0} &
   \g[mMARCO_R10_jaen]{0.0} &
   \g[mMARCO_NDCG_avg]{23.3} &
   \g[mMARCO_NDCG_enes]{30.8} &
   \g[mMARCO_NDCG_enzh]{19.5} &
   \g[mMARCO_NDCG_arru]{20.4} &
   \g[mMARCO_NDCG_jaen]{16.9} \\
 & XLM-R & easy  &
   \g[mMARCO_R10_avg]{16.1} &
   \g[mMARCO_R10_enes]{32.0} &
   \g[mMARCO_R10_enzh]{0.0} &
   \g[mMARCO_R10_arru]{04.0} &
   \g[mMARCO_R10_jaen]{0.0} &
   \g[mMARCO_NDCG_avg]{21.9} &
   \g[mMARCO_NDCG_enes]{29.0} &
   \g[mMARCO_NDCG_enzh]{15.0} &
   \g[mMARCO_NDCG_arru]{17.8} &
   \g[mMARCO_NDCG_jaen]{15.0} \\
 & XLM-R & hard  &
   \g[mMARCO_R10_avg]{83.4} &
   \g[mMARCO_R10_enes]{90.0} &
   \g[mMARCO_R10_enzh]{91.0} &
   \g[mMARCO_R10_arru]{82.0} &
   \g[mMARCO_R10_jaen]{78.0} &
   \g[mMARCO_NDCG_avg]{62.5} &
   \g[mMARCO_NDCG_enes]{69.9} &
   \g[mMARCO_NDCG_enzh]{73.3} &
   \g[mMARCO_NDCG_arru]{58.8} &
   \g[mMARCO_NDCG_jaen]{49.5} \\
\midrule
\multirow{3}{*}{Large-Scale}
 & BM25  & --    &
   \g[LargeScale_R10_avg]{63.5} &
   \g[LargeScale_R10_enes]{76.3} &
   \g[LargeScale_R10_enzh]{07.1} &
   -- & -- &
   \g[LargeScale_NDCG_avg]{57.9} &
   \g[LargeScale_NDCG_enes]{69.7} &
   \g[LargeScale_NDCG_enzh]{06.5} &
   -- & -- \\
 & Nomic & easy  &
   \g[LargeScale_R10_avg]{62.3} &
   \g[LargeScale_R10_enes]{77.0} &
   \g[LargeScale_R10_enzh]{09.0} &
   -- & -- &
   \g[LargeScale_NDCG_avg]{60.7} &
   \g[LargeScale_NDCG_enes]{72.0} &
   \g[LargeScale_NDCG_enzh]{22.7} &
   -- & -- \\
 & XLM-R & easy  &
   \g[LargeScale_R10_avg]{62.0} &
   \g[LargeScale_R10_enes]{77.0} &
   \g[LargeScale_R10_enzh]{09.0} &
   -- & -- &
   \g[LargeScale_NDCG_avg]{57.5} &
   \g[LargeScale_NDCG_enes]{67.6} &
   \g[LargeScale_NDCG_enzh]{22.0} &
   -- & -- \\
\bottomrule
\end{tabular}
}
\caption{Cross-encoder re-ranking with easy and hard negatives over the BM25 baseline. Entries report Recall@10 and nDCG@100 over the top 100 documents per query; improvements over the baseline are shown in \colorbox{green!25}{\textbf{green}} and degradations in \colorbox{red!25}{\textbf{red}}.}

\label{tab:placeholder}
\end{table}

\subsection{Re-ranking Model}

Cross-encoder re-ranking enables finer-grained modelling of interactions between query and document content, at the cost of higher computational load per inference. Across datasets, cross-encoders consistently improve the rank of the relevant document relative to BM25, although the magnitude of the improvement varies across language pairs. The largest gains occur where BM25 performs worst, namely in cross-script or lexically distant pairs such as English-Chinese (en–zh), Arabic-Russian (ar–ru) and Japanese-English (ja–en), where BM25 Recall@10 is close to zero but cross-encoders reach substantially higher values (e.g. CLIRMatrix en–zh improves from 0.0\% to 30.0\%; mMARCO en–zh from 0.0\% to 91.0\% with XLM-R trained on hard negatives). For language pairs with higher lexical overlap such as en–es, improvements are more modest, since BM25 already provides a reasonable initial ranking and the top-100 candidates often contain many plausible but non-relevant documents, which reduces the potential benefit of re-ranking.

Negative sampling emerges as a key factor in determining re-ranking effectiveness. Models trained with hard negatives generally outperform those trained with easy negatives, with XLM-R on mMARCO providing the clearest example: when trained on hard negatives, it reaches an average Recall@10 of 83.4\%, nearly a five-fold improvement over BM25 at 17.3\%, whereas the same architecture trained with easy negatives does not surpass the baseline. Although the available evidence is limited to two datasets and two encoders, this pattern is consistent with prior work that emphasises the importance of hard negatives in retrieval training more broadly, including in cross-lingual settings \cite{Xiong2020_AN, Shen2022_mHFN, Rajapakse2024_NegativeSamplingMultilingual}. In contrast, little effect is observed on the Large-Scale dataset, likely because only easy negatives are available, preventing the cross-encoders from learning to sharpen semantic discrimination.

\begin{table*}[h]
\centering
\begin{tabular}{lccc || ccc}
\hline
& \multicolumn{3}{c||}{\textbf{Average Time}} 
& \multicolumn{3}{c}{\textbf{Average Performance (Recall@100)}} \\
\textbf{Method} 
& CLIRMatrix & mMARCO & Large-Scale 
& CLIRMatrix & mMARCO & Large-Scale \\
\hline
DNN & \textbf{0.388} & \textbf{0.451} & \textbf{0.362} 
    & \textbf{31.9} & \textbf{49.0} &\textbf{64.0} \\
ANN & 0.619 & 0.552 & 0.652 
    & 31.6 & 47.1 & 60.1 \\
\hline
\end{tabular}
\caption{Comparison of average inference time and Recall@100 performance for DNN and ANN retrieval across datasets. Averaged over the five pretrained encoders, over all language pairs in each dataset.}
\label{tab:ann_dnn_combined}
\end{table*}


\paragraph{Retrieval Efficiency and Approximate Indexing.} ANN indexing is typically used to improve retrieval efficiency for embedding-based systems over large corpora. However, in our setting, ANN does \emph{not} yield computational benefits. As shown in Table~\ref{tab:ann_dnn_combined}, ANN search increases average inference time by 40 -- 80\% across datasets compared to direct retrieval, while only reducing Recall@100 by 0.3 -- 3.9\% on average. The document collections are relatively small, ranging from approximately 7k to 26k documents, and exact nearest neighbour retrieval, by comparing the query embedding to all the document embeddings, can be implemented as a single batched cosine-similarity computation between each query embedding and all document embeddings, which is highly optimised and vectorised. By contrast, the approximate setup, based on HNSW indices, requires index access for every query–document language pair and incurs additional overhead from graph traversal and index maintenance. For corpora of this size, these overheads outweigh any pruning benefit from approximate search, resulting in slower retrieval despite only minor reductions in recall.

These observations do not rule out the utility of approximate methods in larger-scale cross-lingual applications, particularly when embedding spaces span many languages and lexical shortcuts are limited. They do indicate, however, that approximate indexing is most effective beyond a certain corpus size. For the benchmarks considered here, approximate search yields negligible improvements in efficiency and slight losses in effectiveness, suggesting that exact dense retrieval remains the more appropriate choice. At larger scales (>1M vectors), ANN has been shown to outperform DNN substantially \cite{johnson2019billion}.

\section{Analysis}
\label{sec:analysis}

The results indicate that the effectiveness of cross-lingual retrieval is shaped not only by the choice of retrieval architecture and fine-tuning strategy, but also by translation decisions and underlying linguistic factors. The following analysis unpacks these influences by examining how retrieval models encode cross-lingual meaning, the role of document translation and supervision, the conditions under which re-ranking is beneficial, and the extent to which linguistic distance affects performance. This provides a qualitative interpretation of the strengths and limitations of current methods beyond aggregate scores.

\paragraph{Influence of Retrieval Models on Performance.} Lexical retrieval is fundamentally constrained by surface-form matching. Translating documents can partially mitigate this limitation, but only when the query language matches the translation language. Dense encoders, in contrast, consistently outperform BM25 on both original and translated corpora because they represent cross-lingual semantics rather than relying solely on shared vocabulary. Contrastive learning on parallel corpora or query–document pairs substantially improves retrieval when applied to base encoders with weak multilingual alignment. For XLM-R, recall increases from baseline values 1.0\%, 4.3\%  and 1.6\% on CLIRMatrix, Large-Scale and mMARCO datasets to 21.2\%, 96.0\% and 77.1\% respectively, bringing performance close to that of Nomic, which is already optimised for cross-lingual retrieval. In contrast, encoders that are already strongly aligned, such as Nomic, exhibit only minor or negligible gains. Taken together, these patterns indicate that semantic alignment between languages and between queries and documents in the embedding space, rather than translation or lexical matching, is the primary driver of cross-lingual retrieval performance.

\paragraph{Impact of Document Translation.}
Because queries cannot reliably be translated at inference time in realistic systems, owing to uncertainty in language identification, variable translation quality and latency constraints, translation is applied only to documents. For BM25, document translation is essential to enable cross-lingual retrieval but is effective only when the query is in the same language as the translated collection. Translating documents into English substantially improves retrieval for English queries by restoring lexical overlap, yet yields little or no benefit for other query languages and fails entirely for cross-script scenarios. 

Embedding-based retrieval behaves differently. Performance is largely unaffected by document translation and can even degrade slightly because of translation-induced noise (Table~\ref{tab:avg_impact_of_translation}). Dense encoders already operate on semantic representations, which reduces dependence on surface form and minimises differences between original and translated conditions. Consequently, while translation supports shallow lexical methods, embedding-based systems derive meaning directly from multilingual representations. Combined with the observations above, this suggests that effective cross-lingual retrieval depends on direct semantic encoding, and that introducing translation into embedding-based pipelines primarily adds computational overhead without improving retrieval quality.

\paragraph{Role of Re-ranking.} Cross-encoder re-ranking can substantially improve the position of the relevant document when the initial candidate set is weak, especially for typologically distant or cross-script language pairs. In these cases, cross-encoders exploit detailed interactions between the query and candidate documents to recover relevant items that first-stage retrieval fails to rank highly. However, the magnitude and consistency of these gains vary across encoders and datasets, and depend strongly on the availability and quality of hard negatives. Models trained with hard negatives generally provide substantial improvements over BM25, whereas training with only easy negatives often fails to surpass the baseline. Re-ranking therefore remains a valuable but non-uniform refinement step. It is most beneficial when strong supervision and additional inference-time computation are available, but its training is complicated by limited lexical overlap and the difficulty of constructing informative cross-lingual negative examples.

\paragraph{Analysis of Retrieval Efficiency.} Efficiency experiments show that approximate nearest neighbour search does not provide a speed advantage at the small to medium corpus scales considered here, with document collections between 7k and 26k items. Index traversal and graph-based overheads in the approximate method outweigh any benefit from pruning, making exact dense nearest neighbour search both faster and more reliable. Approximate indexing thus appears most useful as a scalability mechanism at substantially larger index sizes, rather than as a universal optimisation for cross-lingual retrieval.

\begin{figure*}[h]
    \centering
    
    \begin{subfigure}{0.8\linewidth}
        \centering
        \includegraphics[width=\linewidth]{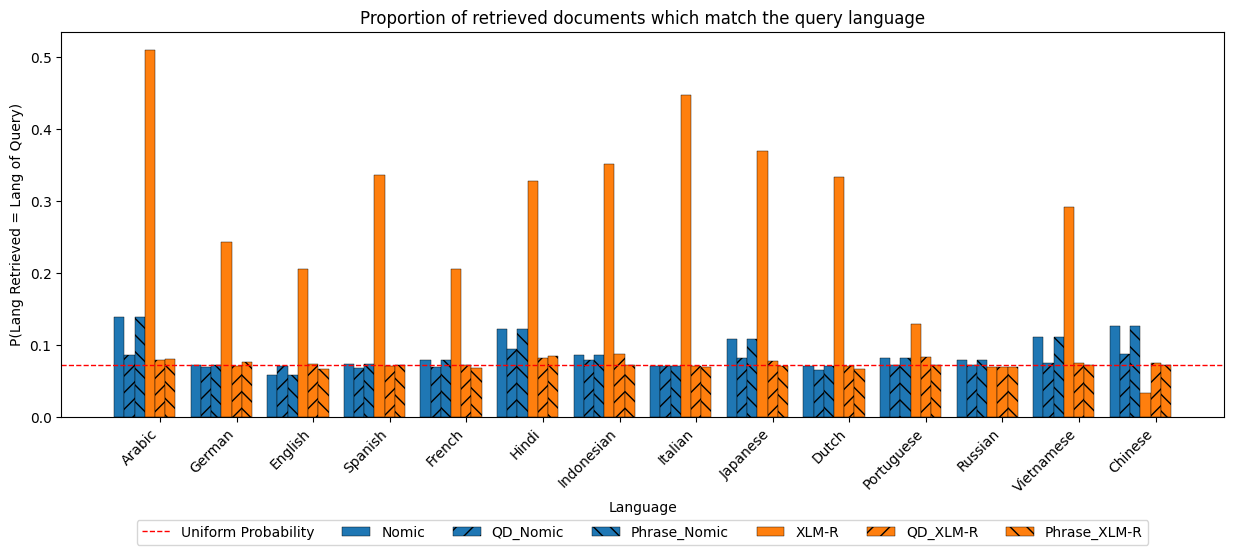}
        \caption{Probability that the model retrieves a document written in the same language as the query. 
        Nomic remains close to uniform, whereas XLM-R without fine-tuning shows strong query-language preference. Bars marked with 
        $(\backslash\backslash)$ and (//) denote phrase-level and query–document contrastive fine-tuning, respectively.}
        \label{fig:q_lang_bias_2}
    \end{subfigure}
    
    \vspace{0.5em}

    \begin{subfigure}{0.8\linewidth}
        \centering
        \includegraphics[width=\linewidth]{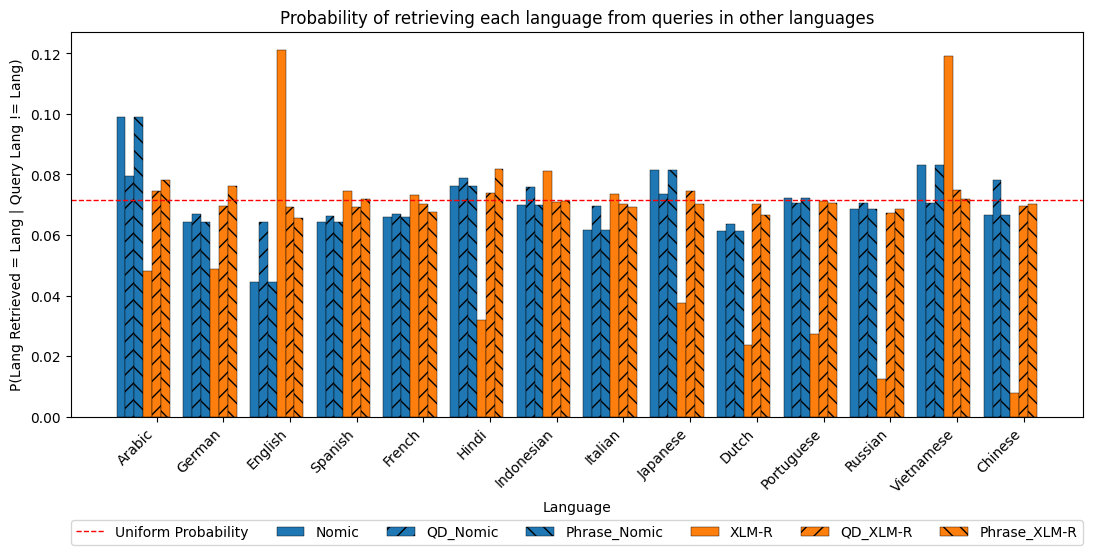}
        \caption{Distribution of retrieved document languages when the correct document is not in the query language. 
        Without fine-tuning, XLM-R displays language and script biases, whereas Nomic retrieves documents more 
        uniformly. Contrastive fine-tuning settings follow the same notation as above.}
        \label{fig:doc_lang_bias_2}
    \end{subfigure}

    \caption{Linguistic bias in document retrieval across Nomic and XLM-R under baseline and contrastive fine-tuning.}
    \label{fig:ling_bias_combined}
\end{figure*}

\paragraph{Linguistic Factors.} To understand why retrieval difficulty varies across language pairs, the quantitative evaluation is complemented by an analysis of how linguistic factors influence accuracy. Two questions are examined: whether linguistic similarity between query and document languages correlates with retrieval quality (Table~\ref{tab:ling_correlation}) and whether models exhibit biases towards particular document languages (Figure ~\ref{fig:ling_bias_combined}).

\begin{table}[h!]
\centering
\begin{tabular}{l l r r r r r}
\toprule
\textbf{Model} & \textbf{Dataset} & \textbf{geographic} & \textbf{syntax} & \textbf{phonology} & \textbf{inventory} & \textbf{genealogical} \\
\midrule

\multirow{3}{*}{BM25} 
    & clirmatrix   & \cc{80.5} & \cc{74.3} & \cc{36.4} & \cc{6.3} & \cc{63.7} \\
    & mmarco       & \cc{48.9} & \cc{70.4} & \cc{41.5} & \cc{38.5} & \cc{54.8} \\
    & large-scale  & \cc{62.9} & \cc{57.2} & \cc{46.7} & \cc{33.6} & \cc{51.3} \\

\midrule

\multirow{3}{*}{XLM-R} 
    & clirmatrix   & \cc{-09.4} & \cc{-28.5} & \cc{-24.4} & \cc{12.8}  & \cc{-21.5} \\
    & mmarco       & \cc{20.2}  & \cc{26.2}  & \cc{16.0}  & \cc{22.3}  & \cc{28.9} \\
    & large-scale  & \cc{31.9}  & \cc{24.5}  & \cc{54.9}  & \cc{23.9}  & \cc{35.7} \\

\midrule

\multirow{3}{*}{Nomic} 
    & clirmatrix   & \cc{53.0} & \cc{65.7} & \cc{30.1} & \cc{-15.3} & \cc{74.4} \\
    & mmarco       & \cc{45.9} & \cc{66.4} & \cc{45.2} & \cc{33.5}  & \cc{71.0} \\
    & large-scale  & \cc{70.0} & \cc{57.4} & \cc{42.9} & \cc{28.8}  & \cc{47.3} \\

\midrule
\multicolumn{2}{l}{\textbf{Average}} 
                     & \cc{44.9} & \cc{46.0} & \cc{32.2} & \cc{20.5} & \cc{45.1} \\
\bottomrule
\end{tabular}
\caption{Spearman correlations between linguistic similarity scores and retrieval performance across language pairs, models, and datasets.}
\label{tab:ling_correlation}

\end{table}

Table \ref{tab:ling_correlation} shows that BM25 retrieval correlates strongly with features associated with typological proximity, particularly geographic, syntactic and genealogical similarity, reflecting its reliance on surface-form overlap. Dense encoders exhibit weaker correlations, indicating an ability to capture semantic alignment beyond lexical similarity. Nonetheless, some structural influence remains: Nomic, the strongest encoder, performs better between linguistically related languages, suggesting that cross-lingual alignment is not fully language agnostic. XLM-R shows very low correlation values, which are likely a consequence of its poor baseline performance rather than evidence of an absence of linguistic effects.

Figures~\ref{fig:q_lang_bias_2} and~\ref{fig:doc_lang_bias_2} reveal fundamental differences in how models encode linguistic information during retrieval. XLM-R disproportionately retrieves documents written in the same language as the query, indicating limited cross-lingual generalisation and semantic abstraction. When the query and document languages diverge, XLM-R exhibits a marked preference for English and surprisingly Vietnamese, while under-retrieving Chinese, Russian, and other languages with distinct scripts. In contrast, Nomic demonstrates a much more uniform retrieval pattern, suggesting stronger cross-lingual alignment.

Critically, we find that contrastive learning, both at the phrase level and query-document level, substantially reduces retrieval bias, bringing XLM-R much closer to the uniform language distribution in the datasets. This suggests that contrastive alignment acts as a corrective mechanism, encouraging models to encode semantic equivalence rather than relying on language-specific features.

Further comparison with additional models (e.g. BM25, LaBSE, multilingual-E5, mmBERT) is provided in Appendix~\ref{app:doc_language_bias}. These results show that language bias is widespread across multilingual encoders and may be driven by their training objectives. Notably, multilingual-E5, despite achieving comparatively strong cross-lingual retrieval performance in Figure~\ref{fig:models_performance}, exhibits very strong query-language retrieval bias, almost on par with BM25. When examining retrieval of documents in non-query languages, models show more similar behaviour; however, most still under-retrieve documents across several languages. This suggests that embeddings for those languages may be encoded in more separated, potentially language-specific clusters.

Overall, the linguistic analysis indicates that language similarity remains a core challenge for purely lexical systems and continues to significantly affect dense retrieval. While strong multilingual encoders substantially mitigate language and script biases, they do not eliminate them entirely. Nonetheless, targeted contrastive learning can further reduce such biases. The disparity between XLM-R and Nomic, as well as improvements observed through fine-tuning, demonstrates that cross-lingual generalisation relies heavily on encoder quality and training objective rather than being an inherent property of dense retrieval architectures.

In summary, the analyses indicate that performance in cross-lingual information retrieval is primarily driven by semantic alignment rather than surface-form similarity or translation effects. Linguistic structure still influences retrieval for weaker models, but its impact diminishes as encoders become more strongly grounded in multilingual semantics. These insights inform the concluding recommendations on the design of robust cross-lingual retrieval systems.

\section{Conclusion}
\label{sec:conclusion}

This work evaluated modelling strategies for CLIR, with a focus on semantic alignment, translation, supervision and the trade-off between efficiency and effectiveness. The study compared five pretrained multilingual encoders, contrastive fine-tuning at multiple levels of granularity, document translation, cross-encoder re-ranking and approximate nearest neighbour search, and examined how linguistic factors influence retrieval behaviour.

The results indicate that effective cross-lingual retrieval depends primarily on models that construct stable, language-agnostic semantic representations. Dense retrieval consistently outperforms lexical and translation-based baselines, confirming that semantic modelling, rather than lexical overlap, is the main driver of performance. Contrastive learning yields substantial gains only for encoders with weak initial alignment, whereas models already optimised for cross-lingual retrieval benefit little from additional supervision. Cross-encoder re-ranking offers inconsistent and often modest improvements, mainly in challenging cross-script or typologically distant language pairs, and its effectiveness is sensitive to the choice of negative documents used during training. Approximate nearest neighbour search does not accelerate inference at the scale of the evaluated benchmarks. Linguistic analysis further shows that structural similarity between languages affects retrieval for weaker models, but this influence diminishes as semantic alignment in the encoder improves. The findings suggest that progress in cross-lingual information retrieval depends on three main factors: (i) strong multilingual pretraining that yields robust language-agnostic embeddings; (ii) targeted contrastive supervision to correct specific alignment deficiencies; and (iii) selective use of translation or re-ranking only when justified by model limitations or typological characteristics of the language pair. Future work could extend these analyses to broader language coverage and larger collections, and explore alignment-aware retrieval strategies, particularly for low-resource and typologically distant languages.

\paragraph{Limitations.} Although this study offers a structured analysis of cross-lingual retrieval across multiple architectures, supervision regimes and linguistic conditions, several limitations constrain the generality of the conclusions. First, the three datasets considered are modest in size, domain coverage and language diversity. This restricts the ability to draw firm conclusions about performance in truly large-scale or highly heterogeneous retrieval scenarios. In particular, the limited number of typologically distant and low-resource languages makes it difficult to fully characterise model behaviour in under-represented linguistic settings. Second, the cross-encoder re-ranking analysis is constrained by the availability of hard negatives in only two datasets, which prevents a systematic assessment of how negative sampling interacts with multilingual semantic alignment. The cross-lingual composition of the cross-encoder training data may also influence how effectively the models learn the task, but this factor is not exhaustively explored here. Third, the efficiency findings are tied to relatively small corpora, so the conclusion that approximate nearest neighbour search does not provide speed benefits is unlikely to hold at larger scales, where index traversal costs are amortised and approximate methods may become advantageous. Finally, the linguistic analyses rely on aggregate typological resources and broad similarity metrics, which may not capture all structural properties that are relevant for cross-lingual retrieval. These limitations highlight the need for larger and more typologically diverse datasets, broader architectural coverage and large-scale evaluations in order to fully characterise the behaviour of cross-lingual retrieval systems.

\bibliographystyle{unsrt}
\bibliography{references}

\clearpage
\appendix

\section{Embedding-Based Retrieval on Translated Corpora Performance}
\label{app:models_performance_translated}

\begin{figure*}[h]
    \centering
    \includegraphics[width=\linewidth]{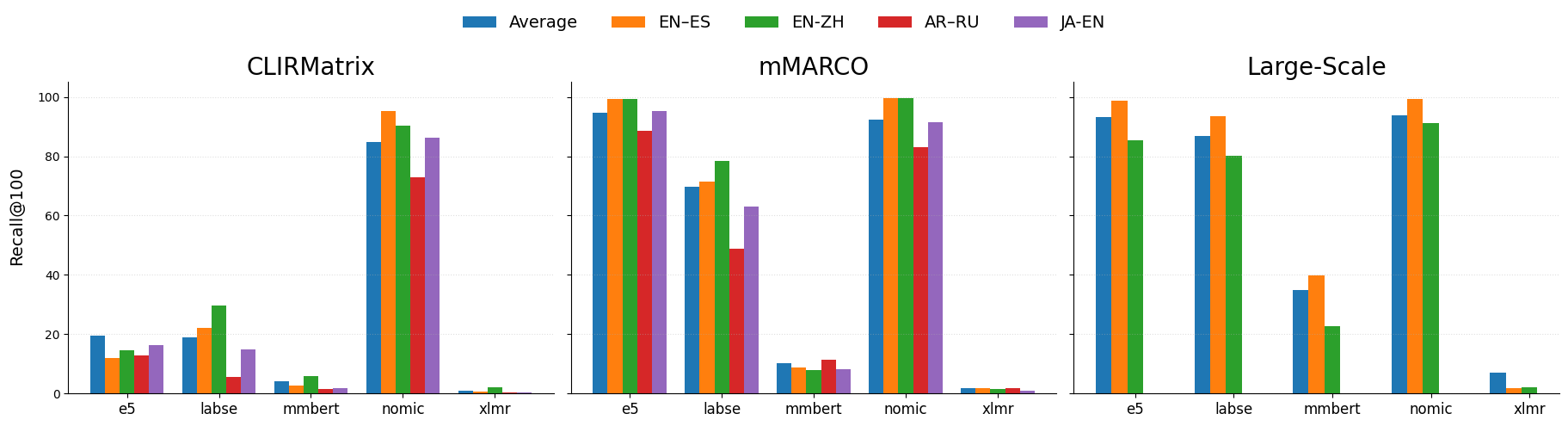}
    \caption{Performance (Recall@100) of five pretrained multilingual encoders using cosine similarity between query and document embeddings on translated document corpora. Results are averaged across all language pairs and shown for selected language pairs.}
    \label{fig:models_performance_translated}
\end{figure*}

\section{Example Query-Document Pairs}
\label{app:dataset_examples}

\begin{table}[h]
\centering
\begin{tabularx}{\linewidth}{l p{0.14\linewidth} X X}
\toprule
\textbf{Dataset} & \textbf{Query} & \textbf{(Translated) Document} & \textbf{Description} \\
\midrule

\textbf{CLIRMatrix} \cite{sun-duh-2020-clirmatrix} 
& Finland
& \textit{Republic of Finland (Finnish: Suomen tasavalta; Swedish: Republiken Finland) is a country in Northern Europe and a member of the European Union since 1995. Finland is bordered by the Baltic Sea…}
& Queries: Wikipedia titles. Documents: Wikipedia articles. Relevance generated via monolingual IR in the query language and propagated across languages using Wikidata interlanguage links. \\

\midrule

\textbf{Large-Scale} \cite{sasaki-etal-2018-cross}
& \textit{Woolmer — is a place in hampshire, england.}
& \textit{Woolmer — Woolmer is a village in East Hampshire, England. It lies 36 km east of Winchester and 71 km southwest of London. In 2010, the village had a population of 550.}
& Queries: first sentences of English Wikipedia articles. Documents: first 200 words of non-English Wikipedia articles (25 languages), with quality filtering. Relevance derived from interlanguage links. \\

\midrule

\textbf{mMARCO} \cite{bonifacio_2022_mmarco}
& where can i find aztec healing clay
& \textit{Secret Aztec healing clay is bentonite clay from Death Valley, California, where it is sun-dried for up to six months at temperatures that sometimes reach 134 degrees. Use it for facials, acne, body wraps, clay baths, foot baths, and chilled clay for knee pads and insect bites…}
& Queries: anonymised Bing search questions. Documents: MS MARCO web passages. Queries and passages machine-translated into multiple languages using Helsinki-NLP models and Google Translate. \\

\bottomrule
\end{tabularx}
\caption{Example queries, documents, and dataset construction details for the three datasets used in this study.}
\label{fig:dataset_examples}
\end{table}

\section{Dataset Preprocessing}
\label{app:preprocessing}

\paragraph{De-duplication and Balancing (mMARCO).}
For the mMARCO dataset, we first deduplicated the translated passages across languages, resulting in a pool of 7{,}433 unique documents. These documents were then evenly distributed across the 14 target languages, assigning exactly one unique passage to each language to ensure balanced representation. For retrieval experiments, the full set of 7{,}433 queries was used for each query language, thereby standardizing the evaluation conditions across all language pairs.

\paragraph{Query Selection.}
For CLIRMatrix and the Large-Scale dataset, we sampled 1{,}000 queries per language pair. In mMARCO, which contains fewer available query–document pairs due to the deduplication step, we used 530–531 queries per language pair (depending on rounding). Sampling was performed uniformly at random from the available query pool for each language combination to avoid potential bias toward any specific topic or language domain.

\section{Retrieval Pipeline Settings}
\label{app:retrievalpipeline}

\subsection{Encoder Model Details}
\label{app:encoder-details}

For all retrieval experiments:

\begin{itemize}[noitemsep]
    \item Tokenization: model-native, maximum 512 tokens  
    \item Pooling: mean pooling over last hidden layer  
    \item Embedding dimension:  
      \begin{itemize}
         \item XLM-R: 768  
         \item LaBSE: 768  
         \item multilingual-E5: 1024  
         \item Nomic: 768  
         \item mmBERT: 768  
      \end{itemize}
    \item Normalization: L2 norm applied before retrieval  
    \item Similarity: cosine similarity  
    \item Caching: all document embeddings precomputed and cached  
\end{itemize}

\subsection{Machine Translation: Evaluation and Settings}
\label{app:translation}

To standardize document representations across languages, all non-English documents from CLIRMatrix and Large-Scale were translated into English. For mMARCO, the provided English passages were used directly. We evaluated four MT systems on 100 randomly sampled documents per language using both semantic adequacy and fluency metrics: \textbf{DeepL} (2024 API), \textbf{NLLB-200}, \textbf{googletrans}, and \textbf{LibreTranslate} (self-hosted v1.5).

\paragraph{Translation configuration.}
Full-scale translation was performed using the \texttt{facebook/nllb-200-distilled-1.3B} checkpoint via the HuggingFace \texttt{translation} pipeline. We used a maximum input length of 1{,}200 tokens, beam search with \texttt{beam=5}, default length penalty, and batch size adapted to available device memory. English and Simple English source documents were copied without modification.

For comparison only, additional translation methods were implemented: DeepL (batch size 10), googletrans with retry-based fallbacks, and LibreTranslate (batch size 20 with per-document fallback). These were used solely for evaluation due to rate limits and instability at scale.

\paragraph{Quality evaluation.}
Semantic adequacy was assessed using COMET-QE (WMT22 COMET-Kiwi) in reference-free mode, comparing each original document with its translation. Fluency was estimated using the negative log-likelihood and perplexity of LLaMA-3.1-8B, computed over translated English text. Across languages, NLLB-200 achieved competitive adequacy and fluency scores (second only to DeepL) while remaining fully scalable and license-permissible. It was therefore selected as the default translation system for all primary experiments.

\subsection{Embedding Setup}
\label{app:embed-setup}

For all cosine-based retrieval experiments, document embeddings were computed once per encoder (batch inference using model-native tokenisation, max 512 tokens), mean-pooled over the final hidden layer, L2-normalized, and cached. Query embeddings were generated on-the-fly using the same configuration. When multiple GPUs were available, documents were distributed across devices and the resulting embeddings concatenated; otherwise, encoding was performed in batches on a single GPU. Retrieval scores were computed via cosine similarity,
\[
\text{sim}(q, d) = \frac{q \cdot d}{\|q\|\|d\|},
\] 
applied over normalised embeddings using matrix multiplication \cite{manning2008introduction}. Documents were ranked by sorting scores in descending order. Optional query-side projection functions were used only in experiments involving alignment intervention.

\subsection{Contrastive Fine-Tuning}
\label{app:contrastive}

We fine-tuned Nomic and XLM-R at three levels: (i) word-level; (ii) phrase-level; and (iii) query–document (QD). All models were fine-tuned end-to-end using a contrastive objective with temperature scaling. Word-level training used a symmetric contrastive loss with dataset-provided negatives, while phrase- and QD-level training used an InfoNCE loss and in-batch negatives. No adapters or projection layers were added on top of the base encoders.

\paragraph{Hyperparameters.}
Table~\ref{tab:contrastive-hparams} summarises the training configuration used for all contrastive setups. Each set up was trained for a maximum of 10 epochs, with post-training best-epoch selection, based on performance on validation data. Batch size was fixed at 16. The learning rate was set to $1\times10^{-5}$ using the Adam optimizer. The loss temperature was $\tau = 0.1$. No scheduler, warmup, weight decay, or gradient clipping was applied.

\begin{table}[h]
\centering
\small
\begin{tabular}{lc}
\toprule
Setting & Value \\
\midrule
Optimizer & Adam \\
Learning rate & 1e--5  \\
Batch size & 16 \\
Loss type & ContrastiveLoss/InfoNCE  \\
Temperature ($\tau$) & 0.1  \\
Max length & 128 (word/phrase), 512 (QD)  \\
Epochs & 10  \\
In-batch negatives & phrase/QD \\
Dataset-provided negatives & word  \\
\bottomrule
\end{tabular}
\caption{Contrastive learning hyperparameters for Nomic and XLM-R.}
\label{tab:contrastive-hparams}
\end{table}

\paragraph{Training data.}
\begin{itemize}[noitemsep]
    \item \textbf{Word-level:} Trained on XL-WiC, Am$^2$iCO, and SenWiCh, retaining only language subsets overlapping with our CLIR datasets. Evaluation was performed using accuracy-based threshold selection.
    \item \textbf{Phrase-level:} Parallel sentence pairs from Tatoeba, sampled up to 1{,}000 examples per language pair. Best checkpoint selected using the sum of A$\to$B and B$\to$A Recall@1.
    \item \textbf{Query–Document:} CLIRMatrix, mMARCO, and Large-Scale training pairs (one model per dataset). In-batch negatives were used. Validation used macro nDCG@10 across language pairs.
\end{itemize}

\paragraph{Implementation.}
All experiments used \texttt{PyTorch} with custom training loops. Checkpoints were selected using validation performance and non-optimal checkpoints were discarded.

\subsection{Re-Ranking (Cross-Encoder) Settings}
\label{app:crossencoder}

\begin{table}[h]
\centering
\small
\begin{tabular}{lc}
\toprule
\textbf{Setting} & \textbf{Value} \\
\midrule
Model architectures & XLM-R-base, Nomic (BERT-style) \\
Max sequence length & 512 (query + document) \\
Per-device train batch size & 8 \\
Per-device eval batch size & 8 \\
Optimizer & AdamW (default in \texttt{sentence-transformers}) \\
Learning rate & 2e--5 \\
LR scheduler & linear with warmup ratio 0.1 \\
Num train epochs & 10 \\
Loss & Binary cross-entropy over sigmoid scores \\
Train/validation split & 90/10, stratified by label \\
Seed & 42 \\
Precision & fp16 on CUDA, full precision otherwise \\
Model selection & best \texttt{eval\_loss}, loaded at end \\
\bottomrule
\end{tabular}
\caption{Cross-encoder fine-tuning hyperparameters.}
\label{tab:crossencoder-hparams}
\end{table}

We fine-tune XLM-R-base and Nomic as cross-encoders that jointly encode each query–document pair and output a scalar relevance score. Training is implemented with \texttt{sentence-transformers} using binary labels ($1$ for relevant, $0$ for non-relevant) and the \texttt{BinaryCrossEntropyLoss} objective.
For each dataset and negative sampling condition (easy vs.\ hard), we train a separate cross-encoder on CSV files containing \texttt{text1} (query), \texttt{text2} (document), and \texttt{label}. The training/validation split is 90/10, stratified by label. Hyperparameters are summarised in Table~\ref{tab:crossencoder-hparams}.

Each positive query–document pair is paired with non-relevant documents according to the difficulty setting. \emph{Easy negatives} are randomly sampled non-relevant documents. \emph{Hard negatives} are derived from dataset-provided candidate rankings and correspond to seemingly relevant but non-match documents.
At inference time, we re-rank the top-100 candidates from the BM25 first-stage ranker. For each query, we form (\textit{query}, \textit{document}) pairs and score them using the trained cross-encoder (batch size 32 during prediction). If the gold document is not present in the BM25 top-100, it is injected by replacing the document at rank 100, ensuring that re-ranking always operates over a candidate set containing the relevant document. Final rankings are obtained by sorting documents by cross-encoder score in descending order, and evaluated using Recall@10 and nDCG@100.

\section{Approximate Nearest Neighbour Search and Efficiency Calculations}
\label{app:ann}

For efficient retrieval, we used HNSW (\texttt{hnswlib}) over L2-normalized document embeddings. The index was created with \texttt{M=16} and \texttt{ef\_construction=200}, storing one entry per document (ID-aligned with embeddings). At inference, queries were encoded using the same encoder and searched with \texttt{ef=50} (or \texttt{max(50, 2$\times$top\_k)} where adaptive). No projection was applied in main experiments.

Cosine similarity was computed as:
\[
\text{sim}(q,d) = 1 - \text{dist}_\text{cosine}(q,d),
\]
and candidates were ranked by descending similarity. Index metadata (model name, embedding dimension, HNSW parameters, and doc IDs) was saved for reproducibility.

\subsection{Latency Measurement Setup}
\label{app:ann-latency}

To compare the efficiency of exact (dense) and approximate (ANN) retrieval, we measured end-to-end inference latency for both methods across all query–document language pairs in each dataset. For each of the five encoder models, retrieval was performed sequentially on each language pair: first using direct cosine similarity over precomputed embeddings (dense retrieval), followed immediately by ANN-based retrieval under identical conditions. Each method produced a timestamp upon completion.

Since different encoder models have varied processing speeds, raw latency values are not directly comparable. To account for this, timestamps were first interleaved in execution order (dense $\rightarrow$ ANN $\rightarrow$ dense $\rightarrow$ ANN, etc.) and min–max normalised to the $[0,1]$ range. Normalised values were then scaled by the number of evaluated language pairs in each dataset (\textit{CLIRMatrix}: 56, \textit{mMARCO}: 196, \textit{Large-Scale}: 26).

For each dataset, we computed the mean latency difference between:
\begin{itemize}[noitemsep]
    \item dense $\rightarrow$ ANN retrieval (i.e., ANN completion time relative to dense), and
    \item ANN $\rightarrow$ dense retrieval (reverse direction in the alternating sequence).
\end{itemize}
The reported latency values correspond to the average of these normalised differences across all models.

\section{Linguistic Similarity Analysis}
\label{app:ling-sim}

To assess whether retrieval performance correlates with linguistic proximity, we computed cosine similarity between query and document languages using typological vectors from \texttt{lang2vec} \citep{littell-etal-2017-uriel}. Feature sets included geographical, syntactic, phonological, phonetic inventory, and family-based attributes; vectors were masked for missing values prior to similarity calculation.

Spearman correlations between language similarity and retrieval \textit{Recall@100} were computed per evaluation file (dataset–model). For comparison across feature sets, we generated a per-file correlation matrix.
Results were aggregated to quantify the influence of linguistic similarity on cross-lingual retrieval performance.

\section{Document Retrieval Language Bias of Evaluated Models}
\label{app:doc_language_bias}

\begin{figure*}[h]
    \centering
    
    \begin{subfigure}{0.8\linewidth}
        \centering
        \includegraphics[width=\linewidth]{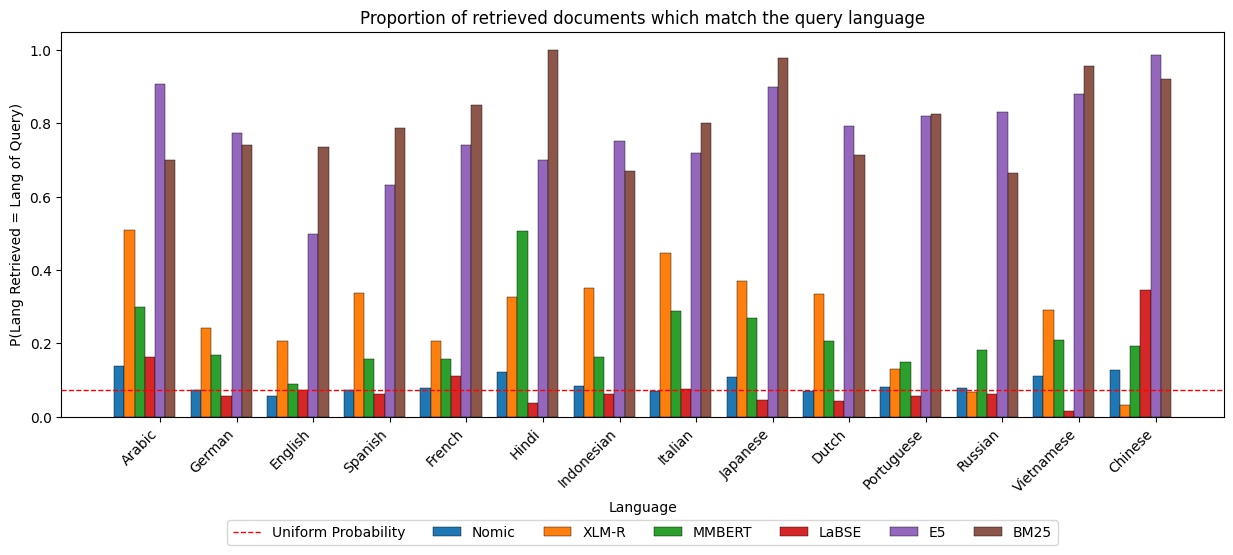}
        \caption{Probability that models retrieve a document written in the same language as the query. 
        BM25 exhibits the strongest bias due to reliance on lexical overlap, while Nomic is the most balanced. 
        Notably, multilingual-E5 also shows a strong preference towards the query language despite strong retrieval performance in Figure~\ref{fig:models_performance}}
        \label{fig:q_lang_bias_models}
    \end{subfigure}

    \vspace{0.5em}

    \begin{subfigure}{0.8\linewidth}
        \centering
        \includegraphics[width=\linewidth]{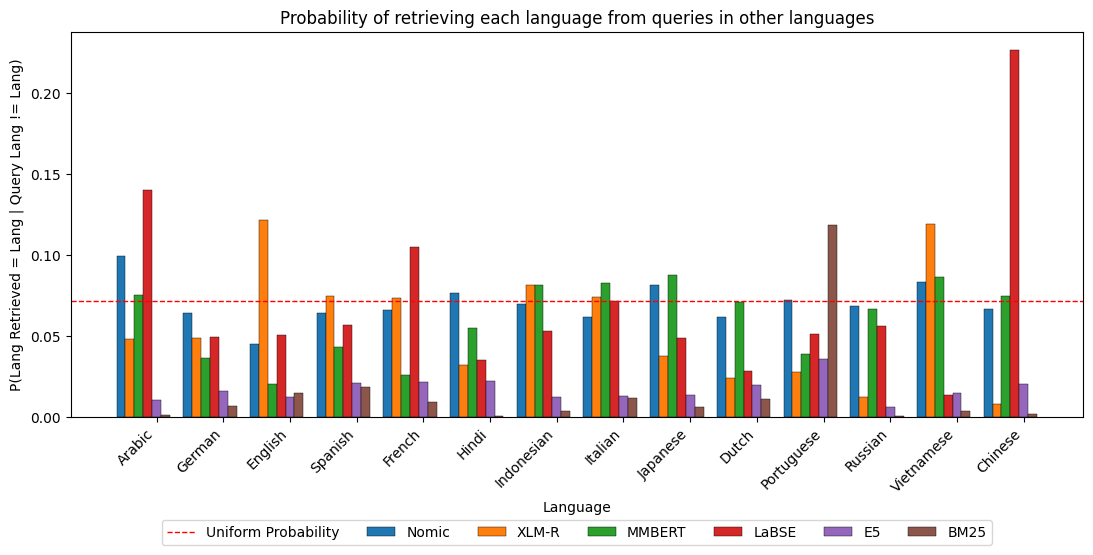}
        \caption{Distribution of retrieved document languages when the relevant document is not in the query language. 
        Most models retrieve non-query languages less frequently than the uniform probability.}
        \label{fig:doc_lang_bias_models}
    \end{subfigure}

    \caption{Comparative linguistic retrieval bias across all evaluated models (Nomic, XLM-R, mmBERT, LaBSE, multilingual-E5 and BM25).}
    \label{fig:ling_bias_combined_appendix}
\end{figure*}

\end{document}